\documentclass[prb]{revtex4}

\usepackage{amsmath}
\usepackage{graphicx}
\usepackage{dcolumn}
\usepackage{bm}
\usepackage[colorlinks,allcolors=black, linktocpage]{hyperref}
\usepackage{caption} 

\DeclareGraphicsExtensions{.png,.pdf}

\newcommand{\Area}{\text{Area}}
\newcommand{\Vol}{\text{Vol}}
\newcommand{\diag}{\text{diag}}

\begin{document}

\preprint{APS/123-QED}

\title{Transport in Superfluid Mixtures}

\author{Michael  Geracie}
 \email{mgeracie@ucdavis.edu}
\affiliation{%
Center for Quantum Mathematics and Physics (QMAP)\\
Department of Physics, University of California, Davis, CA 95616 USA
}%

\date{\today}

\begin{abstract}
We present a general method for constructing effective field theories for non-relativistic superfluids, generalizing the previous approaches of Greiter, Witten, and Wilczek, and Son and Wingate to the case of several superfluids in solution. We investigate transport in mixtures with broken parity and find a parity odd ``Hall drag'' in the presence of independent motion as well as a pinning of mass, charge, and energy to sites of nonzero relative velocity. Both effects have a simple geometric interpretation in terms of the signed volumes and directed areas of various sub-complexes of a ``velocity polyhedron'': the convex hull formed by the endpoints of the velocity vectors of a superfluid mixture. We also provide a simple quasi-one-dimensional model that exhibits non-zero Hall drag.
\end{abstract}

\pacs{Valid PACS appear here}
\maketitle

\section{Introduction}\label{sec:Intro}

Effective field theory (EFT) is a powerful tool for studying the dynamics of dense, strongly coupled matter, where a microscopic description is intractable. In this approach, one writes down an effective action principle for the degrees of freedom that remain at low energies and matches the predictions of this theory against experiment order by order in a momentum expansion. The true predictive power of an EFT lies in the symmetries it possesses, which constrain the space of action principles, so that in practice experimental results may be fit to relatively few parameters.

An EFT for superconductivity was proposed a number of years ago by Greiter, Witten, and Wilczek.\cite{Greiter:1989qb} In this case, the symmetry in question was Galilean invariance, which they imposed by demanding an algebraic relation between the momentum and charge currents
\begin{align}\label{GWWconstraint}
	T^{0i} = \frac{m}{e} j^i .
\end{align}
The physics of this statement is that one expects that for non-relativistic theories, momentum is carried entirely by the transport of matter.

They concluded that to lowest order in a derivative expansion, the most general EFT consistent with this principle is determined by a single function of a single variable
\begin{align}\label{GWWaction}
	S = \int d^{4} x ~ p \left( D_t \varphi - \frac{1}{2m} D_i \varphi D^i \varphi \right) .
\end{align}
In equilibrium, $p(\mu)$ is the thermodynamic pressure as a function of the chemical potential. Here $m$ is the mass of the superconducting order parameter, $\varphi$ is its phase, and $D_\mu \varphi = \partial_\mu \varphi + q A_\mu$. The fact that the low energy dynamics may be entirely characterized by a single function of a single variable demonstrates the power of Galilean symmetry and the utility of the effective action approach.

The theory (\ref{GWWaction}) was pushed to higher order in derivatives by Son and Wingate, who were interested in the next-to-leading order (NLO) physics of the unitary Fermi gas.\cite{Son2006} In this work, the author's introduced and demanded a symmetry called non-relativistic general coordinate invariance, which in particular, implies (\ref{GWWconstraint}). 

However, both of these approaches can be unwieldy. For instance, the constraint (\ref{GWWconstraint}) amounts to a non-linear PDE for the Lagrangian as a function of the fields. While an explicit solution was found to lowest order in derivatives of those fields, this approach is intractable at higher orders as the order of the PDE increases. Non-relativistic general coordinate invariance is a major improvement in this regard and has seen a number of condensend matter applications,\cite{Son:2005tj,Hoyos:2011ez,Son:2013,Andreev:2013qsa,Golkar:2013gqa,Hoyos:2013eha,Hoyos:2014pba,Andreev:2014gia,Gromov:2014gta,Gromov:2014vla,Wu:2014dha,Wu:2014osa,Jensen:2014aia,Jensen:2014ama,Moroz:2014ska,Moroz:2015jla,Fujii:2016mbc,Auzzi:2016lrq}
but often requires a lengthy calculation to confirm invariance in the presence of massive matter. This is particularly true when one lacks, as we shall in this work, a prefered velocity field $v^i$ from which one forms the Galilean invariant combination $\tilde A_\mu$ introduced by Son. It would be advantageous to have a means of writing down manifestly invariant actions for superfluid Goldstones that would remove the need for additional calculation.

More seriously, both methods are intrinsically single constituent in nature: the condition (\ref{GWWconstraint}) relies on this quite explicitly, while non-relativistic general coordinate invariance was motivated as a symmetry of microscopic single constituent actions and does not hold when fields of multiple distinct charge-to-mass ratios are included.
On the other hand, multiconstituent superfluid condensates are of great experimental and theoretical\cite{khalatnikov1957hydrodynamics,khalatnikov1973sound,mineev1974theory,andreev1975three} interest. The most well known example is He3/He4, but experimentally realizable superfluid mixtures have proliferated in recent years due to experimental advances in cold atom physics. The first experimental realization of a superfluid mixture in an atom trap was obtained by Myatt et al. in 1997,\cite{Myatt1997} who condensed the $F=2, m=2$ and $F=1, m=-1$ states of Rb87.  Mixtures can also be created by condensing all the spin states of a single atomic species such as spin-1 Na23.\cite{stenger1998spin}
For an excellent review of weakly coupled BEC mixtures and their experimental realizations we refer the reader to review articles by Kasamatsu, Kawaguchi, and Ueda.\cite{kasamatsu2005,kawaguchi2012spinor} 



We begin in section \ref{sec:Geometry} with an overview of Galilean geometry, which provides an efficient means of writing down EFT's by making the spacetime transformation properties of physical objects manifest. In section \ref{sec:ParityEFT} we apply this to construct the most general parity invariant EFT to lowest order in a derivative expansion. In the single constituent case, this reduces to (\ref{GWWaction}), but allows for a non-dissipative superfluid drag in the general one, an effect originally considered by Andreev and Bashkin.\cite{andreev1975three}

Section \ref{sec:ParityBreakingEFT} extends this analysis to the parity breaking case and contains most of the new results of this paper. In particular, we find a parity odd version of the drag coefficient, which we dub the Hall drag. In two spatial dimensions, this coefficient ``drags'' mass, charge, and energy perpendicular to the relative velocity of two condensates, for instance
\begin{align}
		j^a = c^H \frac{q_1}{m_1} \epsilon^{ab}  (v_{2b} - v_{3b} ) + \cdots .
\end{align}
Galilean invariance also admits a ``pinning'' of mass, charge, and energy to relative velocity, which one may think of as the renormalization of these quantities due to a velocity dependent interaction. For example,
\begin{align}
	j^0 = f \frac{q_1}{m_1} \text{Vol}_{234} + \cdots ,
\end{align}
among other effects.
Here $\Vol_{234}$ is the volume of the 2-simplex that spans the endpoints of the velocity vectors of the three superfluids $2,3,$ and 4. When $\Vol_{234}$ is nonzero, $f$ leads to a buildup of charge proportional to that of the first superfluid. In higher dimensions, these effects have natural geometric interpretations in terms of the signed volumes and directed areas of certain sub-complexes of the convex hull defined by the endpoints of the fluid velocity vectors. For the general expressions, see equations (\ref{hallDragGeneral}) and (\ref{pinningCurrents}).

We conclude in section \ref{sec:Example} with a simple quasi-one-dimensional model which exhibits non-trivial Hall drag and compute $c^H$ in mean field theory.


\section{Galilean Geometry}\label{sec:Geometry}

In this section we recount an efficient method for generating Galilean invariant action principles. This method is essentially an adaptation of pseudo-Riemannian geometry to the Galilean case.\cite{GPR_geometry,GPR_fluids} For massless fields, it reduces to Cartan's treatment of Newtonian gravity. Once this formalism is established, our treatment proceeds straightforwardly along the lines of the relativistic case.\cite{son2002low} Throughout we shall denote spacetime indices by $\mu, \nu, \dots$, with temporal component $t$ and spatial components $i,j,\dots$. Internal Galilean indices in the vector and covector representation will be denoted $A, B, \dots$, with temporal component $0$ and spatial components $a,b,\dots$, while extended indices will be denoted by $I,J,...$. We will regularly pass between spacetime indices and internal Galilean indices using the coframe $e^A_\mu$ and its inverse $e^\mu_A$. In this section we will only give a brief review, a more complete treatment may be found elsewhere.\cite{Geracie:2016bkg} 

\subsection{The Galilean Group}

We begin with the Galilean group $Gal(d)$, where $d$ is the spatial dimension. This is the matrix group
\begin{align}\label{galDef}
	\Lambda^A{}_B = 
	\begin{pmatrix}
		1 & 0 \\
		- k^a & R^a{}_b \\
	\end{pmatrix} ,
\end{align}
where $R^a{}_b$ is a rotation matrix and $k^a$ is the relative velocity between Galilean frames. This is simply the action of Galilean transformations on inertial coordinates $(t ~ x^i)^T$. Note however that $Gal(d)$ will be acting as an internal symmetry throughout, since no natural notion of inertial coordinates exists in the curved case. In this paper we shall follow the approach of Son and many subsequent works in formulating the theory on curved spacetime. This is both a convenient means to make the spacetime symmetries of the theory manifest and an efficient way to encode transport, since a generic spacetime provides the theorist with a suite of knobs to turn to study response.

The velocity $v^A = (1 ~ v^a)^T$ of a particle transforms under this, the vector, representation.
There are however natural objects in non-relativistic physics that are neither Galilean vectors nor singlets. Consider for instance the $(d+2)$-dimensional column vector 
\begin{align}
	p^I = 
	\begin{pmatrix}
		\rho \\
		\rho^a \\
		 - \epsilon
	\end{pmatrix}
\end{align}
where $\rho$ is the mass density, $p^a$ the momentum density, and $\epsilon$ the energy density. The well-known transformation laws for momentum and energy 
\begin{align}
	\rho \rightarrow \rho,
	&&
	p^a \rightarrow - \rho k^a + R^a{}_b p^b  ,
	&& \epsilon \rightarrow \frac 1 2 \rho k_a k^a - k_b R^b{}_a p^a + \epsilon
\end{align}
may be summarized in the matrix
\begin{align}\label{extendedRep}
	p^I \rightarrow \Lambda^I{}_J p^J , 
	&&
	\Lambda^I{}_J =
	\begin{pmatrix}
		1 & 0 & 0 \\
		- k^a & R^a{}_b & 0 \\
		- \frac 1 2 k_c k^c & k_c R^c{}_b & 1 
	\end{pmatrix} .
\end{align}
This forms a representation of $Gal(d)$ called the extended representation. In contrast to the relativistic case then, in which momentum and energy are naturally collected into a $(d+1)$-vector $p^\mu$, in a Galilean covariant theory, mass, energy, and momentum are naturally collected into the $(d+2)$-vector $p^I$.

\subsection{The Extended Derivative}\label{sec:extendedDerivative}

The natural derivative operator on massive non-relativistic fields $\psi$ is not of the form $D_\mu$, familiar from relativistic theories, but is also valued in the extended representation
\begin{align}\label{galDeriv}
	D_I \psi,
\end{align}
as one might guess from the above example by the correspondence $P_i = - i \frac{\partial}{\partial x^i}$, $H = i \frac{\partial}{\partial t}$. Like the energy-momentum $(d+2)$-vector given above, one of the components of this derivative operator is tied to the mass
\begin{align}\label{extendedDerivative}
	D_I \psi =
	\begin{pmatrix}
		D_0 \psi & D_a \psi & i m \psi
	\end{pmatrix} .
\end{align}

For our purposes, the mass $m$ of a non-relativistic field is its representation under $U(1)_M$ transformations\footnote{That this is the same as the kinematic mass - the mass entering the dispersion relation - is fixed by Galilean invariance as can be seen in equation (\ref{schrodAction}). Indeed, Galilean invariance also fixes both to be equal to the gravitational mass.}
\begin{align}\label{U(1)_M}
	\psi \rightarrow e^{i m \alpha} \psi.
\end{align}
Invariance under $U(1)_M$ ensures the existence of a conserved mass current $\rho^\mu$, common to non-relativistic theories. The derivative  operator (\ref{galDeriv}) is both $U(1)_M$ and $Gal(d)$ covariant, and if we tried to do without the final component of (\ref{extendedDerivative}), we would break the later. $U(1)_M$ covariance requires the existence of a mass gauge field, introduced by Duval and K\"unzle into the connection,\cite{Duval:1976ht,DK} which we will return to in section \ref{sec:currents}. Though it will not be important for this work, Newtonian gravity finds its origin in the mass gauge field.

\subsection{Invariant Tensors}

The point of this construction is one may now obtain Galilean invariant action principles straightforwardly by contracting indices with invariant tensors. For the vector representation these are
\begin{align}\label{definingInvariants}
	n_A = 
	\begin{pmatrix}
		1 & 0
	\end{pmatrix},
	&&h^{AB} =
	\begin{pmatrix}
		0 & 0 \\
		0 & \delta^{ab}
	\end{pmatrix}.
\end{align}
The former tensor is called the internal clock-form and provides a non-relativistic theory with an absolute notion of space and time\footnote{In the general case, this decomposition may only be local.} while the latter serves as a spatial metric.

The extended representation admits a higher dimensional version of these invariants
\begin{align}\label{extendedInvariants}
	n_I = 
	\begin{pmatrix}
		1 & 0 & 0
	\end{pmatrix},
	&&g^{IJ} = 
	\begin{pmatrix}
		0 & 0 & 1 \\
		0 & \delta^{ab} & 0 \\
		1 & 0 & 0 
	\end{pmatrix}.
\end{align}
Indeed, the invariants (\ref{definingInvariants}) of the defining representation may be obtained from these by using the projector
\begin{align}
	\Pi^A{}_I = 
	\begin{pmatrix}
		1 & 0 & 0 \\
		0 & \delta^{ab} & 0
	\end{pmatrix} ,
\end{align}
which the reader may check is itself a Galilean invariant. 

Importantly, the extended representation of $Gal(d)$ admits an inverse metric $g^{IJ}$ of Lorentzian signature. We shall denote its inverse by $g_{IJ}$ and use it to raise an lower indices in the usual way. Also note that the form (\ref{extendedDerivative}) of the extended derivative operator may be stated in the invariant way
\begin{align}
	n^I D_I \psi = i m \psi .
\end{align}

Finally, note that since the defining representation of $Gal(d)$ is a subgroup of $SL(d+1)$ and the extended representation of $SL(d+2)$, they also admit the parity and time reversal breaking invariants
\begin{align}
	\epsilon_{A_0 \cdots A_{d}},
	&&\epsilon^{A_0 \cdots A_{d}},
	&&\epsilon^{I_0 \cdots I_{d+1}},
\end{align}
where we have chosen $\epsilon_{0 \cdots d} = \epsilon^{0 \cdots d} = \epsilon^{0 \cdots d+1} = 1$.

As an illustration of this approach, the Schr\"odinger action may be written
\begin{align}\label{schrodAction}
	S = - \frac{1}{2m} \int d^{d+1} x | e |D_I \psi^\dagger D^I \psi = \int d^{d+1} x | e |\left( \frac{i}{2} \psi^\dagger \overset \leftrightarrow D_0 \psi - \frac{\delta^{ab} }{2m} D_a \psi^\dagger D_b \psi \right) ,
\end{align}
where $|e|= \det (e^A_\mu)$ is the volume element.

\subsection{Currents}\label{sec:currents}
In this work we are principally concerned with the currents present in any non-relativistic theory and their response. These currents are most easily defined in a vielbein formalism which we recount here. Since this has been discussed at great length elsewhere, we refer the reader to our references for details.\cite{Geracie:2016bkg}

In this formalism, the geometry is defined by an extended vielbein
\begin{align}
	e^I_\mu = 
	\begin{pmatrix}
		e^A_\mu \\
		a_\mu
	\end{pmatrix}
\end{align}
which contains a spacetime coframe $e^A_\mu = \Pi^A{}_I e^I_\mu$ in its first $d$ components. The final component is the mass gauge field alluded to in section \ref{sec:extendedDerivative} and transforms as such $a_\mu \rightarrow a_\mu + \nabla_\mu \alpha$ under (\ref{U(1)_M}). It is the star of Newton-Cartan geometry and is the manner in which it encodes Newtonian gravitational effects,\cite{Duval:1976ht,DK} reducing to the familiar Newtonian gravitational potential after fixing an appropriate Galilean frame. 
However, for our purposes, its principal role is to serve as a source for mass current. The extended derivative operator is determined by this data, and under a variation $\delta e^I_\mu$ and $\delta A_\mu$ we have
\begin{align}\label{derivativeVariation}
	\delta D_I \psi = - \Pi^\mu{}_I \left( \delta e^J_\mu D_J \psi + i q \psi \delta A_\mu \right) ,
\end{align}
where we have also chosen to couple to a background electromagnetic potential $A_\mu$ with charge $q$.

The currents are then defined as\footnote{We are specializing to the case of vanishing spin current which will be relevant for our discussion. A general treatment may be found elsewhere.\cite{Geracie:2016bkg} It would be interesting to extend the analysis of this paper to magnetized superfluid mixtures, such as the ferromagnetic or polar phases of condensed Rb87 or Na23 by incorporating spin.}
\begin{align}
	\delta S = \int \left( - \tau^\mu{}_I \delta e^I_\mu + j^\mu \delta A_\mu \right) .
\end{align}
The extended-valued tensor $\tau^\mu{}_I$ encodes the flow of mass, energy, and stress. In a particular Galilean frame its components have the following interpretation\footnote{Note that $- n_A \tau^A{}_I$ is the mass-momentum-energy current we considered above (\ref{extendedRep}).}
\begin{align}\label{stressEnergyComponents}
	\tau^\mu{}_I =
	\begin{pmatrix}
		\epsilon & - \rho^i & - \rho \\
		\epsilon^i & - T^i{}_a & - \rho_a 
	\end{pmatrix},
\end{align}
where $\epsilon$ is the energy density, $\rho$ the mass density, $\epsilon^i$ and $\rho^i$ their associated currents, and $T^i{}_a$ the stress tensor. When all indices are converted to the same type using the vielbein, the stress tensor is symmetric on-shell by Ward identities.\cite{GPR_improv}

\section{Parity Invariant Superfluid Mixtures}\label{sec:ParityEFT}

We now turn to the problem studied by Greiter, Witten, and Wilczek, that of a single component charged superfluid at $T=0$. The only degree of freedom remaining at low temperatures is a single superfluid phase $\varphi$. It is a simple matter to determine how the covariant derivative operator acts on a phase $\psi = e^{- i \varphi} | \psi |$
\begin{align}
	D_I \varphi = 
	\begin{pmatrix}
		D_0 \varphi &
		D_a \varphi &
		- m
	\end{pmatrix},
\end{align}
where
\begin{align}
	D_A \varphi = e^\mu_A (\partial_\mu \varphi + q A_\mu + m a_\mu ) .
\end{align}
Here $e^\mu_A$ is the inverse vielbein, $A_\mu$ the electromagnetic vector potential, and $a_\mu$ the mass gauge field. In what follows we will often work with the ``extended velocity'' of the condensate and its projection
\begin{align}
	v_I = - \frac{1}{m} D_I \varphi =
	\begin{pmatrix}
		- \frac 1 m D_0 \varphi &
		- \frac 1 m D_a \varphi &
		1
	\end{pmatrix},
	&&v^A = \Pi^{AI} v_I
	= \begin{pmatrix}
	1 \\
	- \frac 1 m D^a \varphi
	\end{pmatrix}.
\end{align}

\subsection{Effective Action}

Due to the shift symmetry $\varphi \rightarrow \varphi + c$, the phase must always enter the action with derivatives. The natural Galilean covariant action to lowest order in derivatives is then
\begin{align}\label{oneComponentAction}
	S &= \int d^{d+1} x | e | p \left(- \frac{1}{2m} D_I \varphi D^I \varphi \right) \nonumber \\
		&= \int  d^{d+1} x | e | p \left( D_0 \varphi - \frac{1}{2m}D_a \varphi D^a \varphi \right) ,
\end{align}
with an arbitrary function $p$, generalizing (\ref{GWWaction}) to curved space.
It would be interesting to carry out this analysis to higher orders, generalizing the work of Son and Wingate\cite{Son2006} beyond NLO. For this work, we are considering backgrounds that are small deviations from flat spacetime with no applied electromagnetic field. That is, our power counting scheme is
\begin{align}
	D_I = O (\epsilon),
	&&A_\mu = \mathcal O (\epsilon),
	&&a_\mu = \mathcal O (\epsilon)
	&&e^A_\mu = \delta^A{}_\mu + \mathcal O ( \epsilon ),
\end{align}
so that additional, background dependent terms do not enter at lowest order. The superfluid velocity $D_I \varphi$ may however be large.

It now should be clear how to generalize to arbitrary superfluid mixtures. The natural set of Galilean invariants we may form from a collection of phases $\varphi_i$ with masses $m_i$ and charges $q_i$ is
\begin{align}
	\mu_{ij} &= - D_I \varphi_i D^I \varphi_j \nonumber \\
		&= m_j D_0 \varphi_i + m_i D_0 \varphi_j - D_a \varphi_i D^a \varphi_j .
\end{align}
There is also a single additional set of invariants which must be considered for completeness, however, they do not lead to any distinct effects, so we relegate consideration of them to appendix \ref{app:SuperfluidDrag}.
The lowest order EFT is then
\begin{align}
	S = \int d^{d+1}x | e | p ( \mu_{ij} ).
\end{align}

What transport does this encode? Using the variation (\ref{derivativeVariation}) as well as $\delta |e| = | e | \Pi^\mu{}_I \delta e^I_\mu$, we find that
\begin{align}\label{covariantCurrents}
	n^\mu_i &= n_i v^\mu_i + \frac 1 2 \rho \sum\nolimits' c^d_{ij} \frac {1}{m_i}  ( v^\mu_i - v^\mu_j ) , \nonumber \\
	j^\mu &= \sum q_i n_i v^\mu_i + \frac 1 2 \rho \sum\nolimits' c^d_{ij} \left( \frac{q_i}{m_i} - \frac{q_j}{m_j}\right) ( v^\mu_i - v^\mu_j ) , \nonumber \\
	\tau^\mu{}_I &= - p \Pi^\mu{}_I - \sum m_i n_i v^\mu_i v_{jI}  
		- \frac 1 2 \rho \sum\nolimits' c^d_{ij} (v^\mu_i - v^\nu_j) (v_{iI} - v_{jI}) .
\end{align}
Primed summations denote a sum over all independent components of the tensor structures appearing in the summand, here, $i < j$.
$n^\mu_i$ is the Noether current generated by the symmetry $\varphi_i \rightarrow \varphi_i + c$. $n_i$ and $\rho$ are then interpreted as the number density of the $i$th species and the total mass density respectively. $c^d_{ij}$ is a phenomenon unique to the multiconstituent case which we will have more to say on in a moment.

In the above, we have defined
\begin{align}\label{dragDef}
	n_i &= \sum_j N_{ij}, 
	&&\rho = \sum_i m_i n_i , 
	&&c^d_{ij} = - \frac{2}{\rho} m_i N_{ij} .
\end{align}
We refer to the matrix
\begin{align}\label{numberMatrix}
	N_{ij} = S_{ij} m_j, 
	&&\text{where}
	&&S_{ij} = 2 p_{ij},
	&&\delta p = \sum_{ij} p_{ij} \delta \mu_{ij} 
\end{align}
as the number matrix.\footnote{Note that since the sum in (\ref{numberMatrix}) is unconstrained, there is some double counting. For instance $p_{ij} = \frac 1 2 \frac{\partial p}{\partial \mu_{ij}}$ when $i \neq j$.} Though perhaps notational overkill at this stage, these definitions will prove convenient when we compute effective masses in appendix \ref{app:EffectiveMass}.

To understand this better, let's write down these formulas in the more familiar component form of (\ref{stressEnergyComponents})
\begin{align}\label{nonCovariantCurrents}
	n_i^A &=
	\begin{pmatrix}
		n_i \\
		n_i v^a_i + \frac 1 2 \rho  \sum_j c^d_{ij} \frac{1}{m_i} ( v^a_i -  v^a_j ) 
	\end{pmatrix}, \nonumber \\
	j^A &=
	\begin{pmatrix}
		\sum q_i n_i \\
		\sum q_i n_i v^a_i + \frac 1 2 \rho  \sum' c^d_{ij} \left( \frac{q_i}{m_i} - \frac{q_j}{m_j}\right) (v^a_i - v^a_j )
	\end{pmatrix}, \nonumber \\
	\rho^A &=
	\begin{pmatrix}
		\sum m_i n_i \\
		\sum m_i n_i v^a_i
	\end{pmatrix}, \nonumber \\
	\epsilon^A &=
	\begin{pmatrix}
		\sum \mu_i n_i - p \\
		\sum \mu_i n_i v^a_i + \frac 1 2 \rho  \sum' c^d_{ij} \left( \frac{\mu_i}{m_i} - \frac{\mu_j}{m_j}\right) ( v^a_i - v^a_j ) 
	\end{pmatrix}, \nonumber \\
	T^{ab} &= p \delta^{ab} + \sum m_i n_i v^a_i v^b_i 
	 +  \frac 1 2 \rho \sum\nolimits' c^d_{ij} (v^a_i - v^a_j)(v^b_i - v^b_j) ,
\end{align}
where we have defined the energy per particle $\mu_i = D_0 \varphi_i$ (this is the same as the chemical potential in the homogeneous case).

\subsection{Superfluid Drag}\label{sec:SuperfluidDrag}

This decomposition has an obvious interpretation: the fluid of density $n_i$ carries mass $m_i$, charge $q_i$, and energy $\mu_i$ per particle in the direction $v^a_i$. The fluid has pressure $p$ and the standard kinetic contribution to the stress is fixed by Galilean invariance. The coefficients $c^d_{ij}$ are the {\it drag coefficients}. In the presence of a relative velocity of the $i$th and $j$th superfluids, they lead to a force per unit area
\begin{align}
	\frac{dF}{dA} = \frac 1 2 \rho c^d_{ij} (\Delta v_{ij})^2
\end{align}
directed along the relative velocity vector, the standard definition of the drag coefficient in hydrodynamics.\cite{landau1987fluid} 

As originally observed by Mineev,\cite{mineev1974theory} steady-state configurations with independent motion exist even in the presence of superfluid drag. This is immediately seen from the equations of motion
\begin{align}
	\nabla_\mu n^\mu_i = 0,
\end{align}
which are trivially satisfied with static and homogeneous densities $n_i$ and velocities $v^\mu_i$, regardless of their relative orientations. In particular, superfluid drag does not introduce dissipation in a superfluid mixture with independent motions. This holds regardless of microscopic dynamics and in particular applies as well in the presence of the parity odd generalizations to drag that we will consider in the rest of this paper.

In the presence of superfluid drag, the mass, energy, and charge currents are not simple weighted sums of the number current, but there is rather additional transport induced by the mutual  interactions of the superfluids. This phenomenon was incorporated into the hydrodynamic description of condensate mixtures by Andreev and Bashkin,\cite{andreev1975three} who anticipated the effect on the following physical grounds. In the presence of interactions between two atomic species in mixture, the first species is transformed into a quasi-particle excitation of effective mass $m^\star_1$ greater than its bare mass $m_1$. A flow of the the first superfluid then must carry with it some mass of the second, even if the second has no number current. Roughly, mass, charge, and energy are ``dragged'' along the direction of relative velocity. In this discussion we have taken the definition of the velocity to be parallel to the number current, whereas in the rest of this paper we have defined the velocity to be parallel to the momentum of a given superfluid component $v^a_i = - \frac{1}{m} D^a \varphi_i$. One may pass from one description to the other by a simple redefinition of variables.

The sound velocities may be obtained from the equations of motion $\nabla_\mu n^\mu_i = 0$ at the linearized level
\begin{align}
	2 \sum_{jkl} S_{ij,kl} \partial_t^2 \varphi_l = \sum_j S_{ij} \nabla^2 \varphi_j ,
\end{align}
with $S_{ij}$ defined in (\ref{numberMatrix}) and $\delta S_{ij} = \sum S_{ij,kl} \delta \mu_{kl}$. This result will not be altered by considerations in subsequent sections except to alter the expression for $S_{ij}$ in equation (\ref{sDef}). In the single component case (\ref{oneComponentAction}) one may check that this reduces to the familiar result
\begin{align}
	\partial_t^2 \varphi = c_s^2 \nabla^2 \varphi ,
	&&c_s = \sqrt{\frac{\partial p}{\partial \rho}}.
\end{align}

\section{Parity Breaking Superfluid Mixtures}\label{sec:ParityBreakingEFT}

The principle results of this paper concern parity and time-reversal breaking transport. This would be relevant in the presence of a background magnetic field, or say, in mixtures of chiral molecules.\footnote{We thank Dam Son for the later suggestion.} We find the symmetries admit two types of parity odd transport that can only be achieved in the presence of superfluids in mixture. The first is a parity odd version of the drag coefficient just discussed, which we dub the Hall drag, while the second pins charge, mass, and energy to relative velocity, in addition to other effects. The number of superfluids required to realize each possibility depends on the dimensionality. A simple microscopic model that realizes the Hall drag in $1+1$ dimensions is given in section \ref{sec:Example}.

\subsection{The Hall Drag}\label{sec:HallDrag}
The first example we consider is a parity odd version of the drag coefficients considered in section \ref{sec:SuperfluidDrag}. For simplicity we begin in $2+1$ dimensions in a tripartite mixture $\varphi_1, \varphi_2, \varphi_3$. We may then form an additional $P$ and $T$ breaking scalar
\begin{align}
	\lambda &= - \epsilon_{ABC} D^A \varphi_1 D^B \varphi_2 D^C \varphi_3 \nonumber \\
		&= m_1 \epsilon^{ab} D_a \varphi_2 D_b \varphi_3 + m_2 \epsilon^{ab} D_a \varphi_3 D_b \varphi_1 
		 + m_3 \epsilon^{ab} D_a \varphi_1 D_b \varphi_2 ,
\end{align}
where $D^A \varphi_i = \Pi^{AI} D_I \varphi_i = \begin{pmatrix} - m_i & D^a \varphi_i \end{pmatrix}^T$. Note that this invariant requires the presence of at least three condensates, and the generalization to the case of more than three condensates should be clear. As in the parity invariant case, there is an additional set of scalars that can be constructed that leads to the same effect, which for simplicity of presentation we relegate to appendix \ref{app:Parity}.

The pressure is then a function of the new variable $\lambda$ in addition to $\mu_{ij}$
\begin{align}
	S = \int d^3 x |e| p ( \mu_{ij} , \lambda ) 
\end{align}
and the currents are those found in (\ref{covariantCurrents}), plus
\begin{align}
	n^\mu_1 &= c^H \epsilon^\mu{}_{\nu \lambda} \frac{1}{m_1} v^\nu_2 v^\nu_3\nonumber \\
	j^\mu &= c^H \epsilon^\mu{}_{\nu \lambda} \left( \frac{m_1}{q_1} v^\nu_2 v^\lambda_3 + \frac{m_2}{q_2} v^\nu_3 v^\lambda_1 + \frac{m_3}{q_3} v^\nu_1 v^\lambda_2 \right) , \nonumber \\
	\tau^\mu{}_I &= - c^H \epsilon^\mu{}_{\nu \lambda} ( v_{1I} v^\nu_2 v^\lambda_3 +  v_{2I} v^\nu_3 v^\lambda_1 +  v_{3I} v^\nu_1 v^\lambda_2)  ,
\end{align}
and cyclic permutations for the other $n^\mu_i$. The general formula with an arbitrary number of condensates may be found in equation (\ref{hallCurrents}). Here
\begin{align}
	c^H &= m_1 m_2 m_3 \frac{\partial p}{\partial \lambda}.
\end{align}
These may be computed using the variations (\ref{oddVariations}).

In a fixed ``lab frame'' (\ref{stressEnergyComponents}), these read
\begin{align}
	&n_1^A =
	\begin{pmatrix}
		0 \\
		c^H \frac{1}{m_1} \epsilon^{ab}(v_{2b} -  v_{3b} )
	\end{pmatrix}, \nonumber \\
	&j^A =
	\begin{pmatrix}
		0 \\
		c^H \frac{q_1}{m_1} \epsilon^{ab}  (v_{2b} - v_{3b} ) + \cdots 
	\end{pmatrix}, \nonumber \\ 
	&\rho^A = 0, \nonumber \\
	&\epsilon^A =
	\begin{pmatrix}
		0 \\
		c^H  \frac{\mu_1}{m_1}  \epsilon^{ab} (v_{2b} - v_{3b}) + \cdots 
	\end{pmatrix}, \nonumber \\
	&T^{ab} = c^H v^{(a}_1 \epsilon^{b )c} (v_{2c} - v_{3c}) + \cdots ,
\end{align}
where the ellipses indicate cyclic permutations. We have symmetrized the stress by hand, since we know that local rotation invariance implies on-shell symmetry of the stress tensor.\cite{GPR_improv}

We see that, much like the drag $c^d$, $c^H$ leads to stresses when two fluid components are in relative motion and induces currents that are not the weighted sum of the densities times velocity.
These currents are ``dragged perpendicular'' to relative velocity rather than along it, but are otherwise precisely of the same form as the standard drag currents, so we will refer to $c^H$ as the {\it Hall drag}.

\begin{figure}[t]
		\centering
		\includegraphics[width=.2\linewidth]{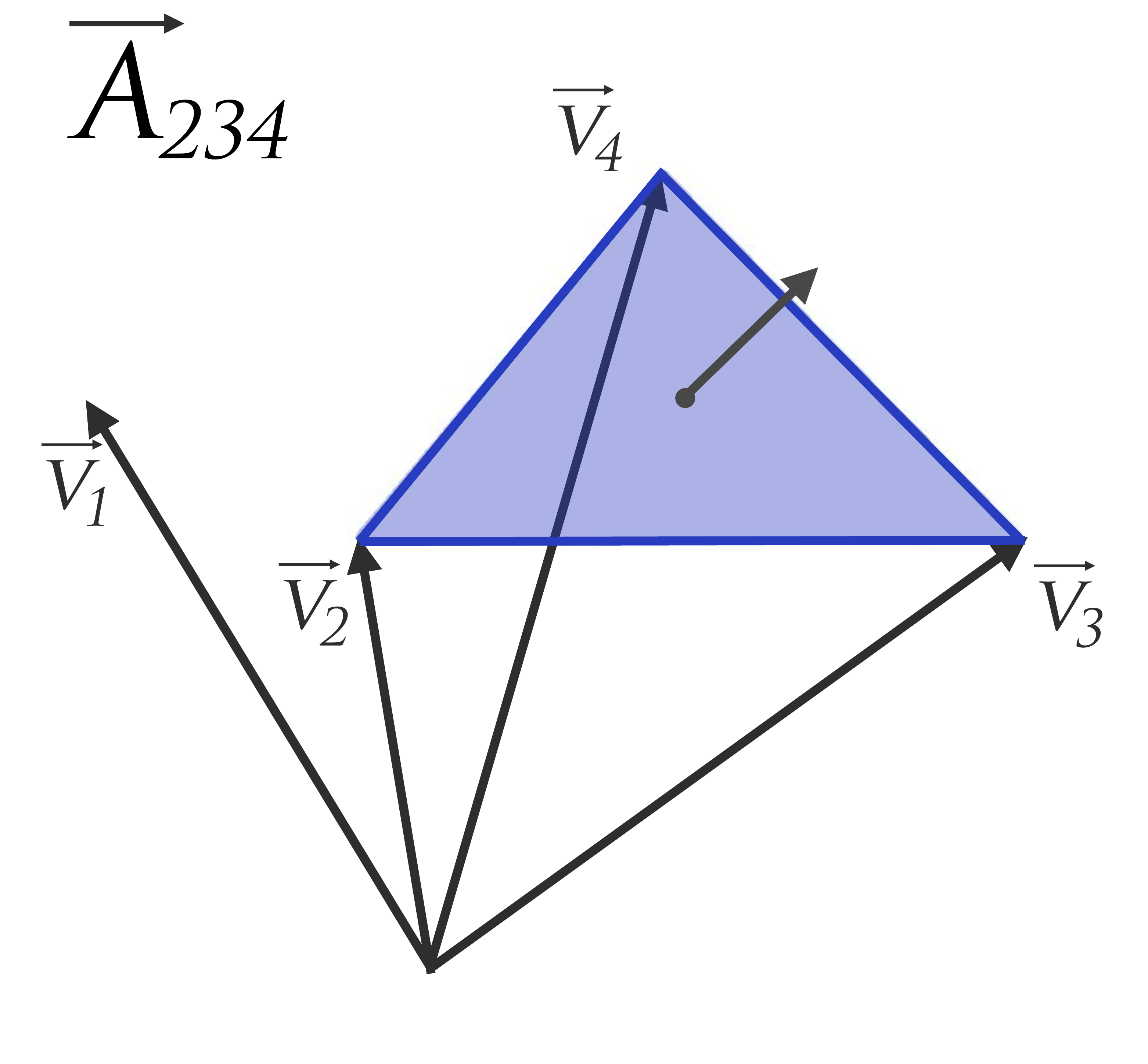}
		\qquad \includegraphics[width=.2\linewidth]{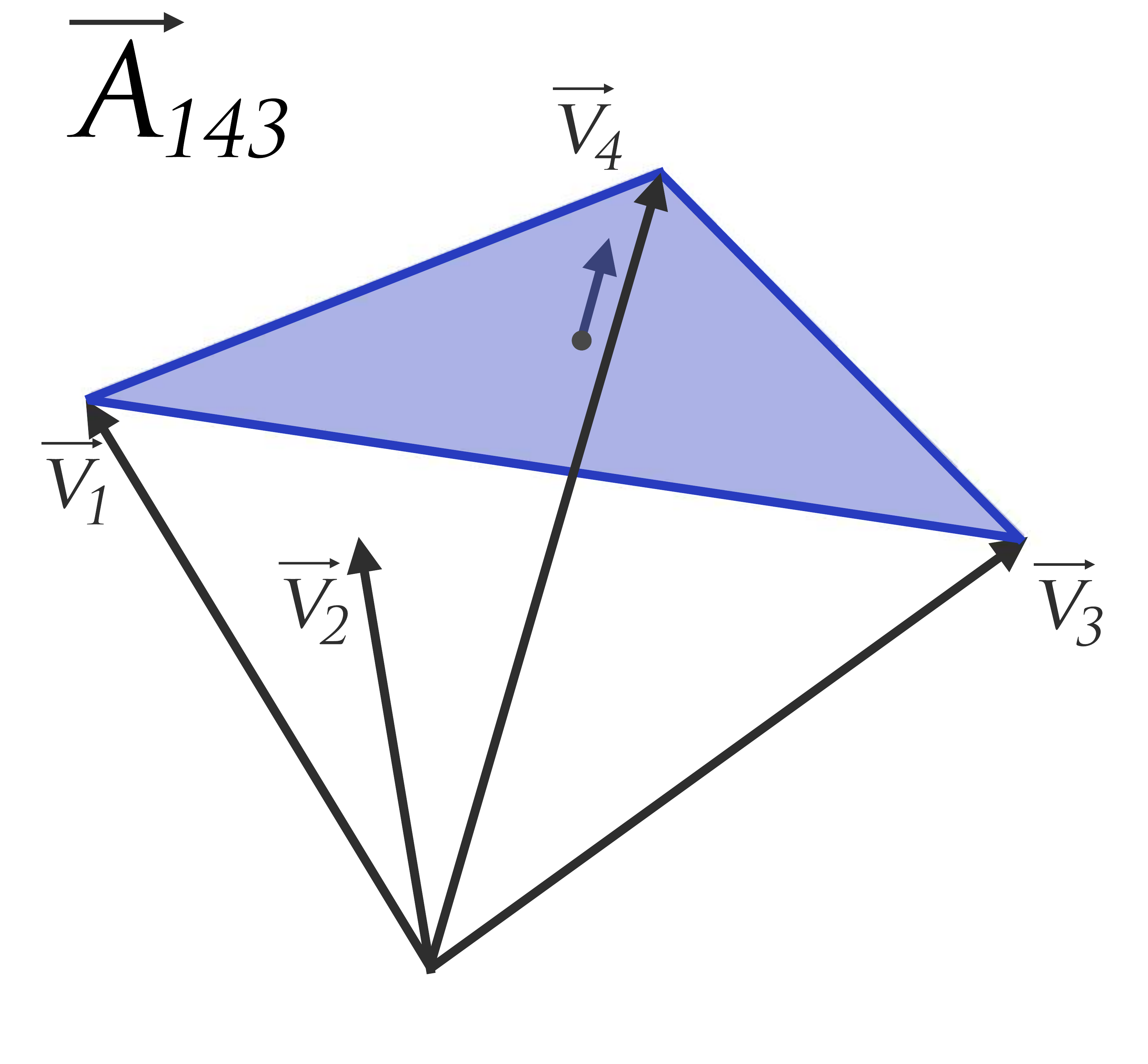}
		\qquad \includegraphics[width=.2\linewidth]{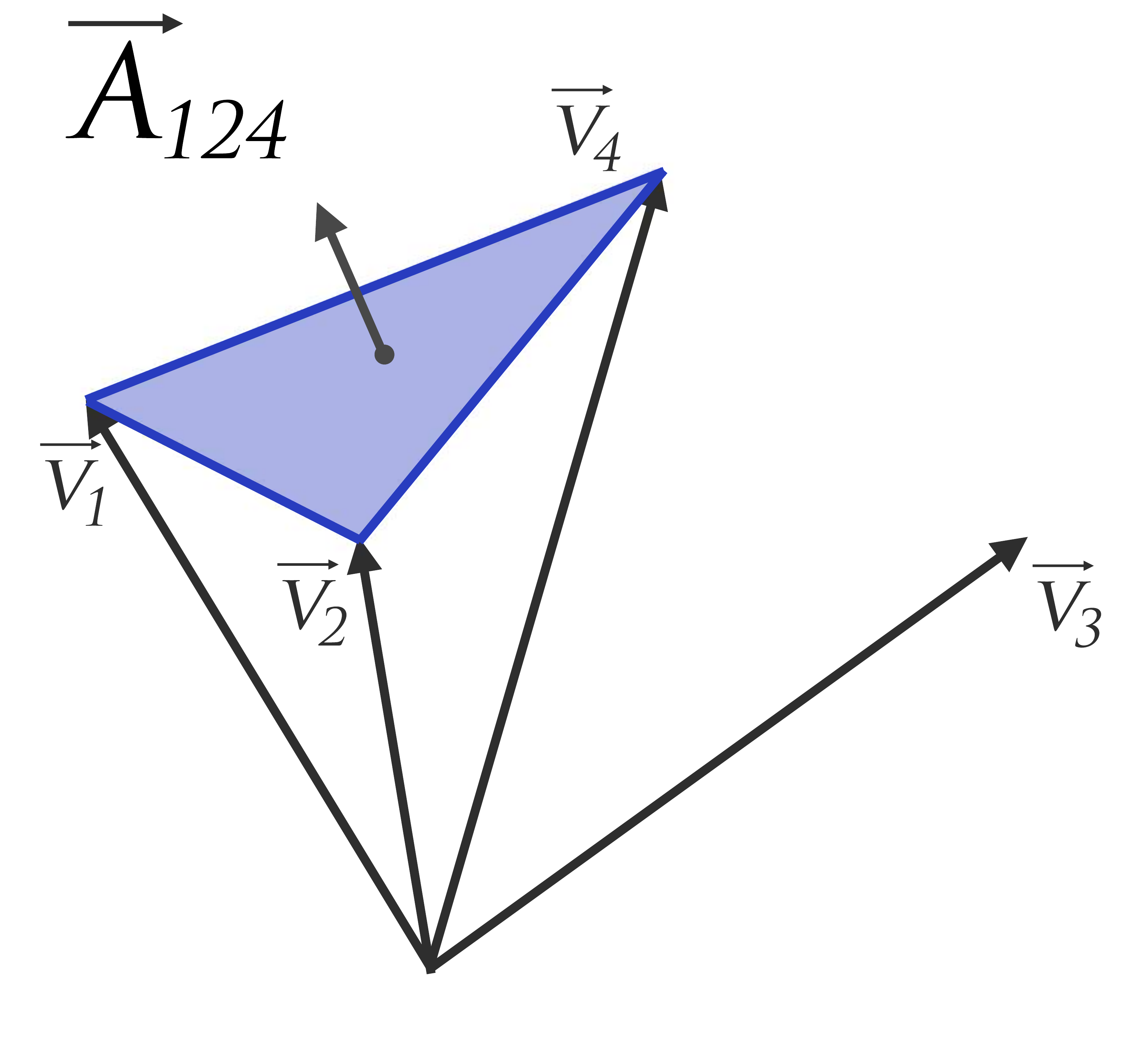}
		\qquad \includegraphics[width=.2\linewidth]{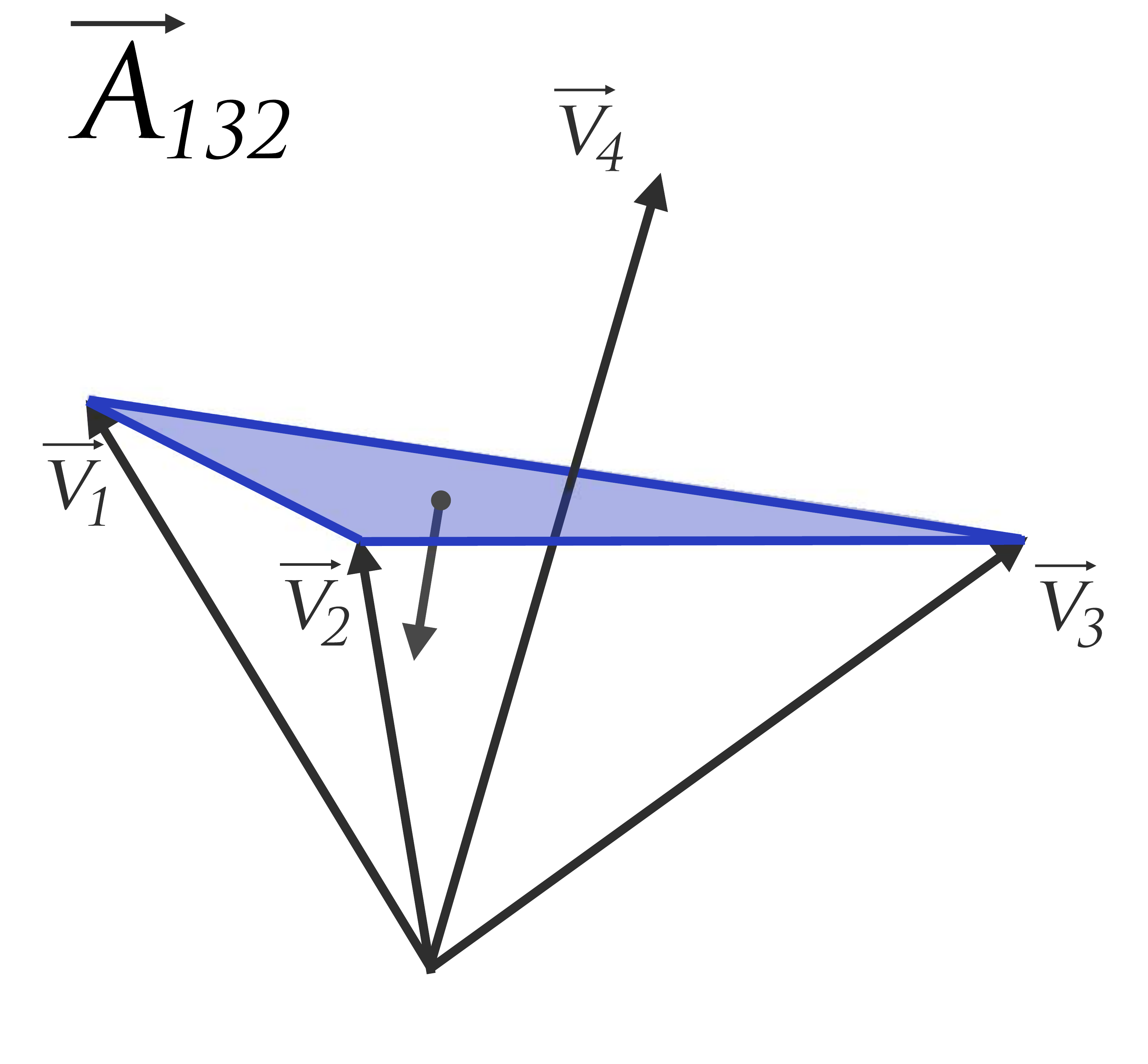}
\caption{The directed areas contributing to Hall drag in $3+1$ dimensions.}
\label{fig:Areas}
\end{figure}

This behavior generalizes naturally to higher dimensions and any number of condensates. For each condensate $i$, one forms all $(d-1)$-simplices whose corners are determined by the velocity vectors of any $d$ other condensates $i_1 , \dots , i_d$. The Hall drag then drags the $i$th condensate along the directed area $\Area^a_{i_1 \cdots i_d}$ of this simplex
\begin{align}\label{hallDragGeneral}
	&n_i^A =
	\begin{pmatrix}
		0 \\
		\sum\nolimits' c^H_{i i_1\cdots i_d} \frac{1}{m_{i}} \text{Area}^a_{i_1 \cdots i_d}
	\end{pmatrix} , \nonumber \\
	&j^A =
	\begin{pmatrix}
		0 \\
		\sum\nolimits' c^H_{i i_1\cdots i_d} \frac{q_{i}}{m_{i}} \text{Area}^a_{i_1 \cdots i_d}
	\end{pmatrix} , \nonumber \\
	&\rho^A = 0 , \nonumber \\
	&\epsilon^A =
	\begin{pmatrix}
		0 \\
		 \sum\nolimits' c^H_{i i_1 \cdots i_d} \frac{\mu_{i}}{m_{i}}\text{Area}^a_{i_1 \cdots i_d}
	\end{pmatrix} , \nonumber \\
	&T^{ab} = \sum \nolimits ' c^H_{i i_1 \cdots i_d} v^{(a}_{i} \text{Area}^{b)}_{i_1 \cdots i_d} ,
\end{align}
where
\begin{gather}
	\text{Area}^a_{i_1 \cdots i_d} = \frac{1}{(d-1)!} \epsilon^{a a_1 \cdots a_{d-1} } (\Delta v_{i_1 i_2})_{a_1}  \cdots (\Delta v_{i_{d-1} i_d})_{a_{d-1}} , \nonumber \\
	\text{and} 
	\qquad \qquad c^H_{i_0 \cdots i_d} = (d+1)!(d-1)! m_{i_0} \cdots m_{i_d} p_{i_0 \cdots i_d}.
\end{gather}
with $\delta p = \sum p_{i_0 \cdots i_d} \delta \lambda_{i_0 \cdots i_d} $.
This procedure is illustrated in figure \ref{fig:Areas}. In this picture, Galilean invariance is the statement that this proceedure is independent of the choice of origin.

Note that there is an interesting interplay between dimensionality and the number of overlapping condensates necessary to realize this effect: it only exists when the number of condensates exceeds the spatial dimensionality of the system. For instance in the $(3+1)$-dimensional case relevant in most experiments, one requires at least 4.

\subsection{Pinning Charge to Relative Velocity}\label{sec:velocityPinning}
There is a single additional set of parity odd scalars one may construct. Again specializing to $2+1$ dimensions, but now in the presence of $4$ condensates, this is
\begin{align}
	\xi &= \epsilon^{IJKL} D_I \varphi_1 D_J \varphi_2 D_K \varphi_3 D_L \varphi_4 \nonumber \\
		&= ( m_1 D_0 \varphi_4 - m_4 D_0 \varphi_1 ) \epsilon^{ab} D_a \varphi_2 D_b \varphi_3 + \cdots ,
\end{align}
and the effective action is
\begin{align}
	S = \int d^3 x | e | p ( \mu_{ij} , \lambda_{ijk} , \xi ) .
\end{align}
As we shall see, this invariant pins charge, energy, and particle number to sites of relative velocity, in addition to other effects. In other words, $\xi$ encodes a velocity dependent interaction that alters the effective mass, charge, and chemical potential so that these densities are not simple weighted sums of $n_i$ with their bare values $m_i, q_i, \mu_i$.

In covariant form, this induces transport
\begin{align}
	&n_1^\mu =    \frac{f}{ 2}\epsilon^{\mu}{}_{ I J K} \frac{1}{m_1} v^I_2 v^J_3 v^K_4 , \nonumber \\
	&j^\mu =    \frac{f}{2}\epsilon^{\mu}{}_{ I J K} \frac{q_1}{m_1} v_2^{I} v_3^{J} v_4^{K} + \cdots , \nonumber \\
	&\tau^\mu{}_I = - \frac{f}{2}\epsilon^\mu{}_{JKL} v_{1 I} v_2^J v_3^K v_4^L + \cdots,
\end{align}
and cyclic permutations to obtain the other $n^\mu_i$'s, where $\epsilon^{\mu JKL} = \Pi^\mu{}_I \epsilon^{IJKL}$ and
\begin{align}
	f = - 2 m_1 m_2 m_3 m_4 \frac{\partial p}{\partial \xi} .
\end{align}
For the general formula, see (\ref{fCovariant}).
Expressing this in the lab frame (\ref{stressEnergyComponents}), we find
\begin{align}
	n_1^A &= 
	\begin{pmatrix}
		f \frac{1}{m_1} \text{Vol}_{234} \\
		\frac{f}{2} \left( \frac{1}{m_1} \frac{\mu_2}{m_2} - \frac{1}{m_2} \frac{\mu_1}{m_1} \right) \epsilon^{ab} (v_{3b} - v_{4b} ) + \cdots
	\end{pmatrix}, \nonumber \\
	j^A &= 
	\begin{pmatrix}
		f \frac{q_1}{m_1} \text{Vol}_{234} + \cdots\\
		\frac{f}{2} \left( \frac{q_1}{m_1} \frac{\mu_2}{m_2} - \frac{q_2}{m_2} \frac{\mu_1}{m_1} \right)\epsilon^{ab} (v_{3b} - v_{4b} ) + \cdots
	\end{pmatrix}, \nonumber \\
	\rho^A &= 0, \nonumber \\
	\epsilon^A &= 
	\begin{pmatrix}
		f \frac{\mu_1}{m_1} \text{Vol}_{234} + \cdots \\
		0
	\end{pmatrix}, \nonumber \\
	T^{ab} &= - \frac{1}{2} f \left( \frac{\mu_1}{m_1} v^{(a}_2 - \frac{\mu_2}{m_2} v^{(a}_1 \right) \epsilon^{b)c} (v_{3c} - v_{4c} ) + \cdots ,
\end{align}
where
\begin{align}
	\Vol_{234} = \frac 1 2 \epsilon_{ab} \Delta v_{23}^a \Delta v_{34}^b.
\end{align}
We see that $f$ induces charge and number transport perpendicular to relative velocity as the Hall drag does, however, the magnitude of the effect is proportional to the bare energies per particle $\mu_i$.

Moreover, as previously mentioned, $f$ pins additional charge, energy, and particle number to sites of relative velocity.\footnote{In the perhaps more natural picture where density is defined to be the zero component of $n_i^\mu$, one would say it pins additional charge, energy, and mass to sites of relative velocity.} The amount is proportional to the signed volume $\Vol_{ijk}$ of the 2-simplex formed by connecting the endpoints of any three velocity vectors. The greater the relative velocities, the stronger the interaction and the more pronounced the effect.

\begin{figure}[t]
		\centering
		\includegraphics[width=.2\linewidth]{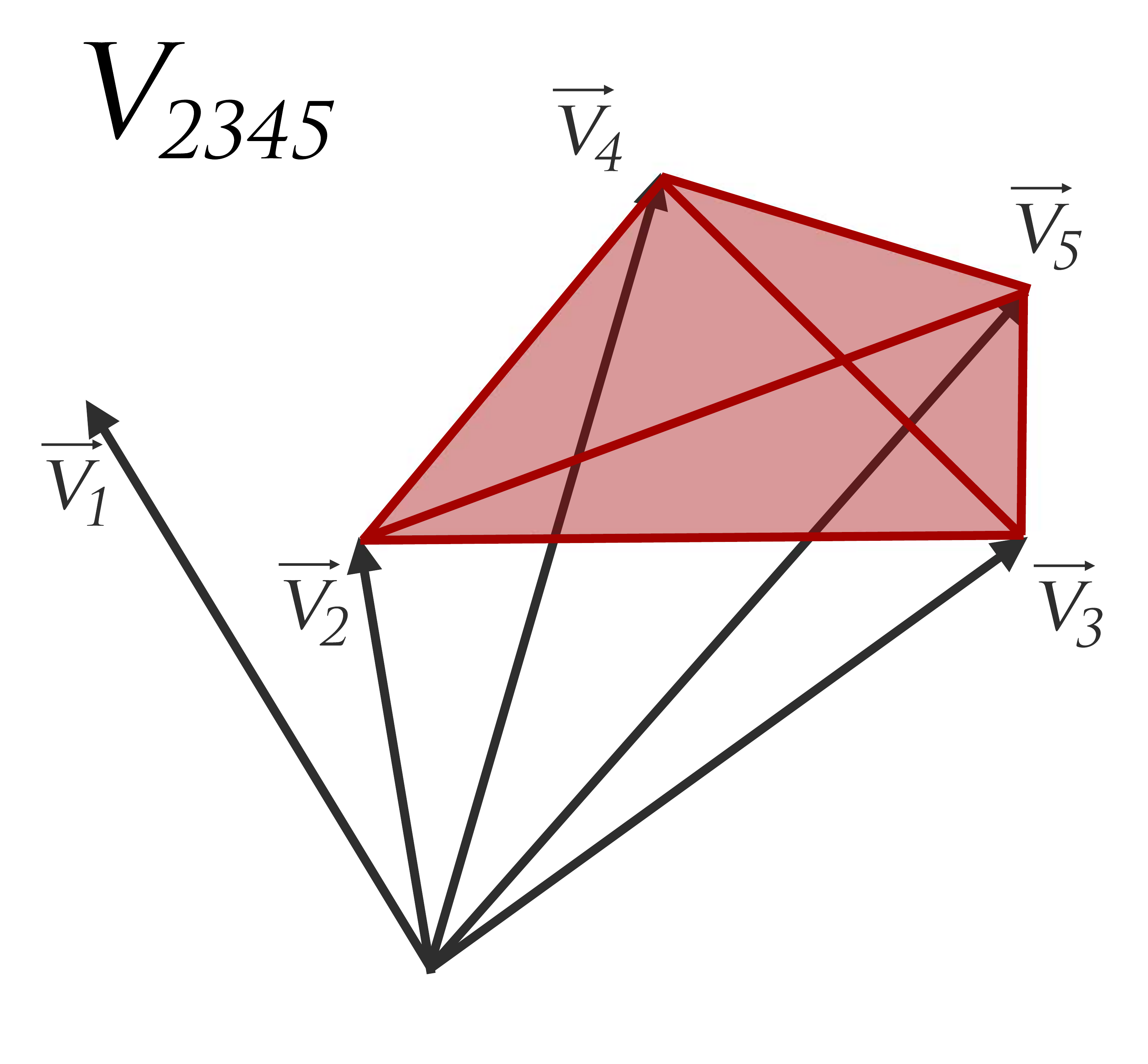}
		\qquad\qquad \includegraphics[width=.2\linewidth]{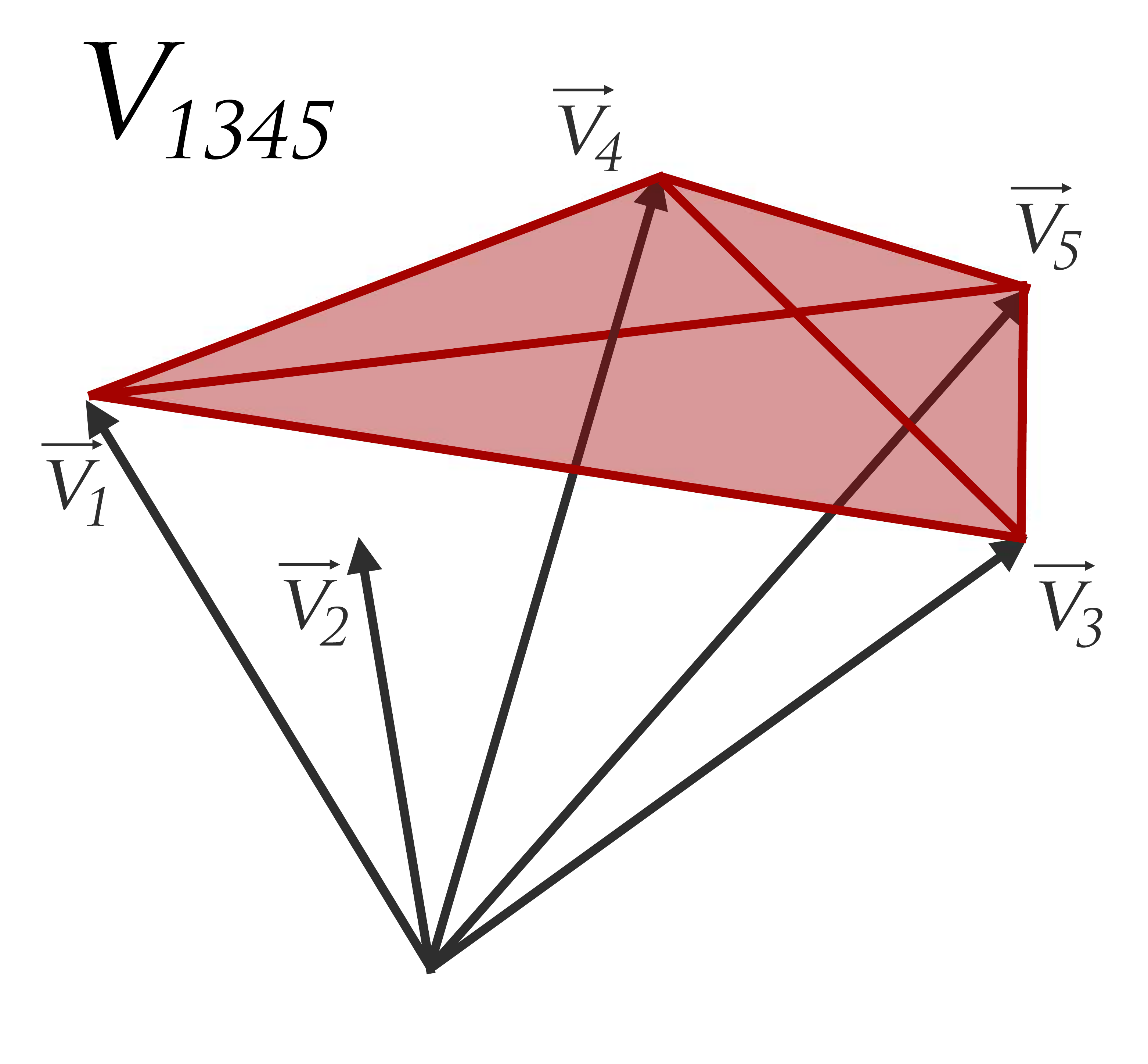}
		\qquad\qquad \includegraphics[width=.2\linewidth]{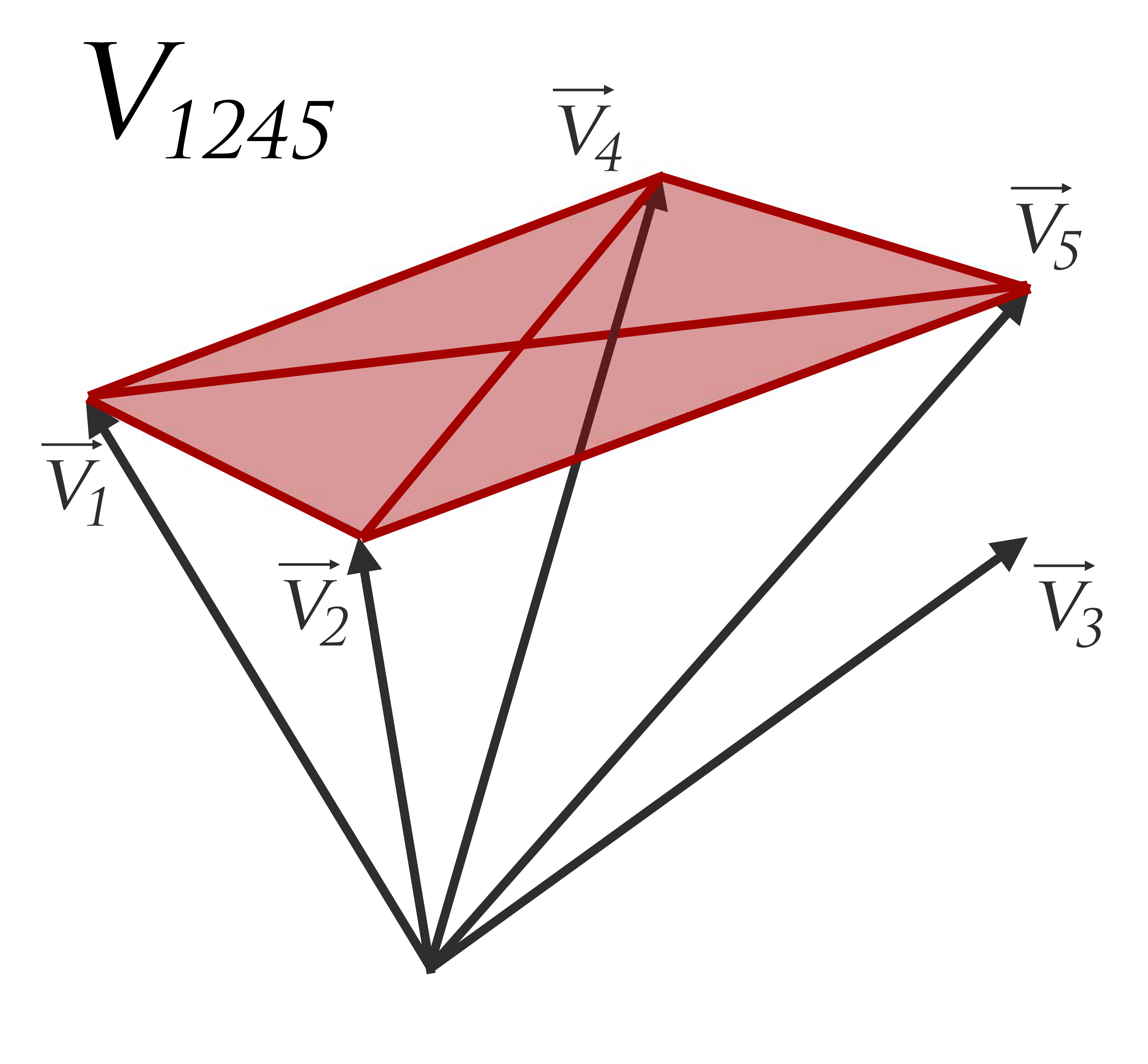}
\caption{Some of the signed volumes contributing to (\ref{pinningCurrents}) in $3+1$ dimensions.}
\label{fig:Volumes}
\end{figure}

As with the Hall drag, these formulas generalize naturally to any dimension and involve the signed volumes and directed areas of various simplices. To find the amount of fluid $i$ pinned, form all $d$-simplices whose corners are determined by the velocity vectors of any $d+1$ other condensates $i_1 , \dots , i_d$. $f$ pins an amount of $i$ proportional to the signed volume $\Vol_{i_1 \cdots i_{d+1}}$ of this simplex
\begin{align}
	\Vol_{i_1 \cdots i_{d+1}}
		&= \frac{1}{d!} \epsilon_{a_1 \cdots a_d }\Delta v^{a_1}_{i_1 i_2} \cdots \Delta v^{a_d}_{i_d i_{d+1}} .
\end{align}
This procedure is pictured in figure \ref{fig:Volumes}.

\begin{figure}[b]
		\centering
		\includegraphics[width=.2\linewidth]{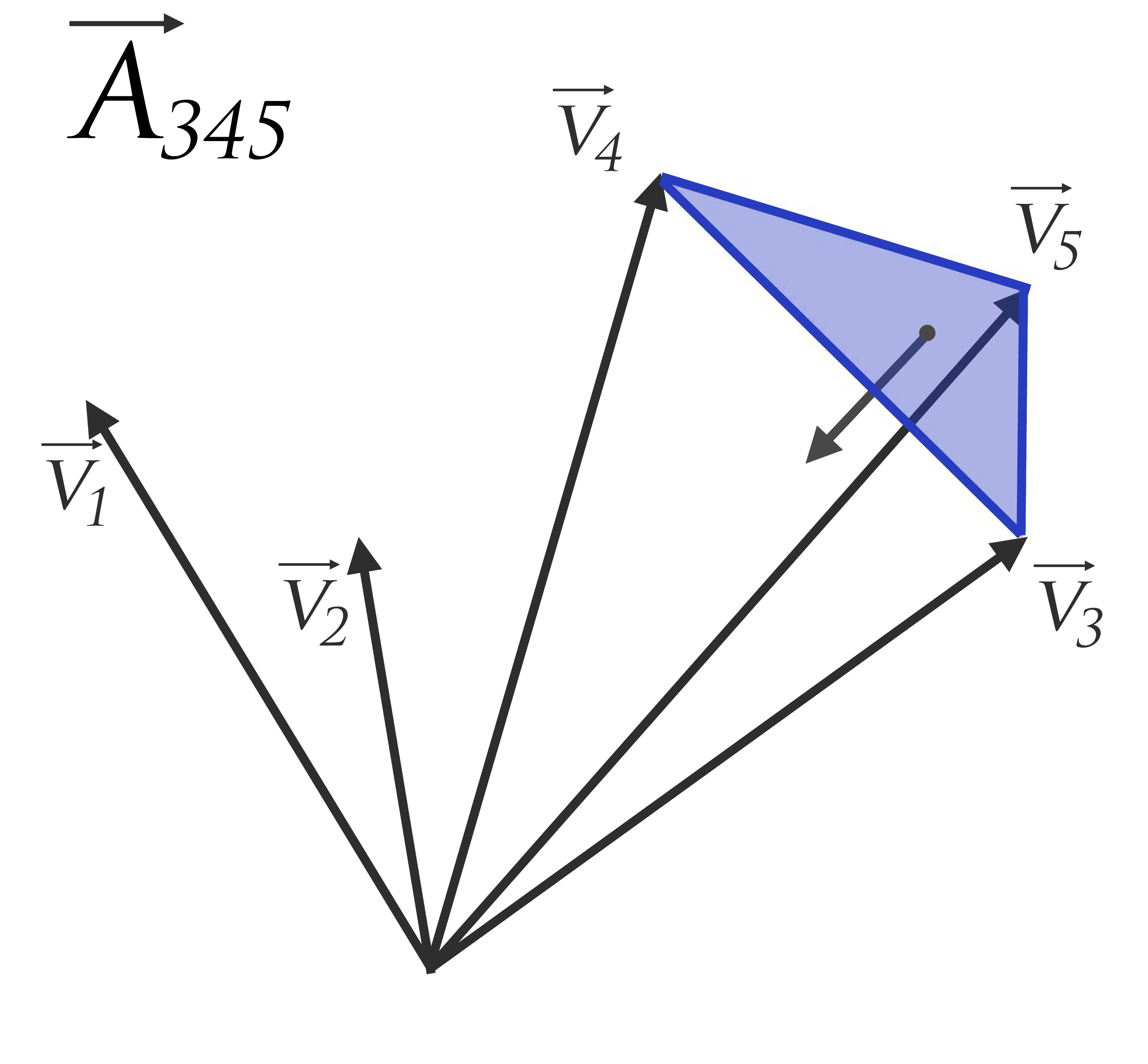}
		\qquad\qquad \includegraphics[width=.2\linewidth]{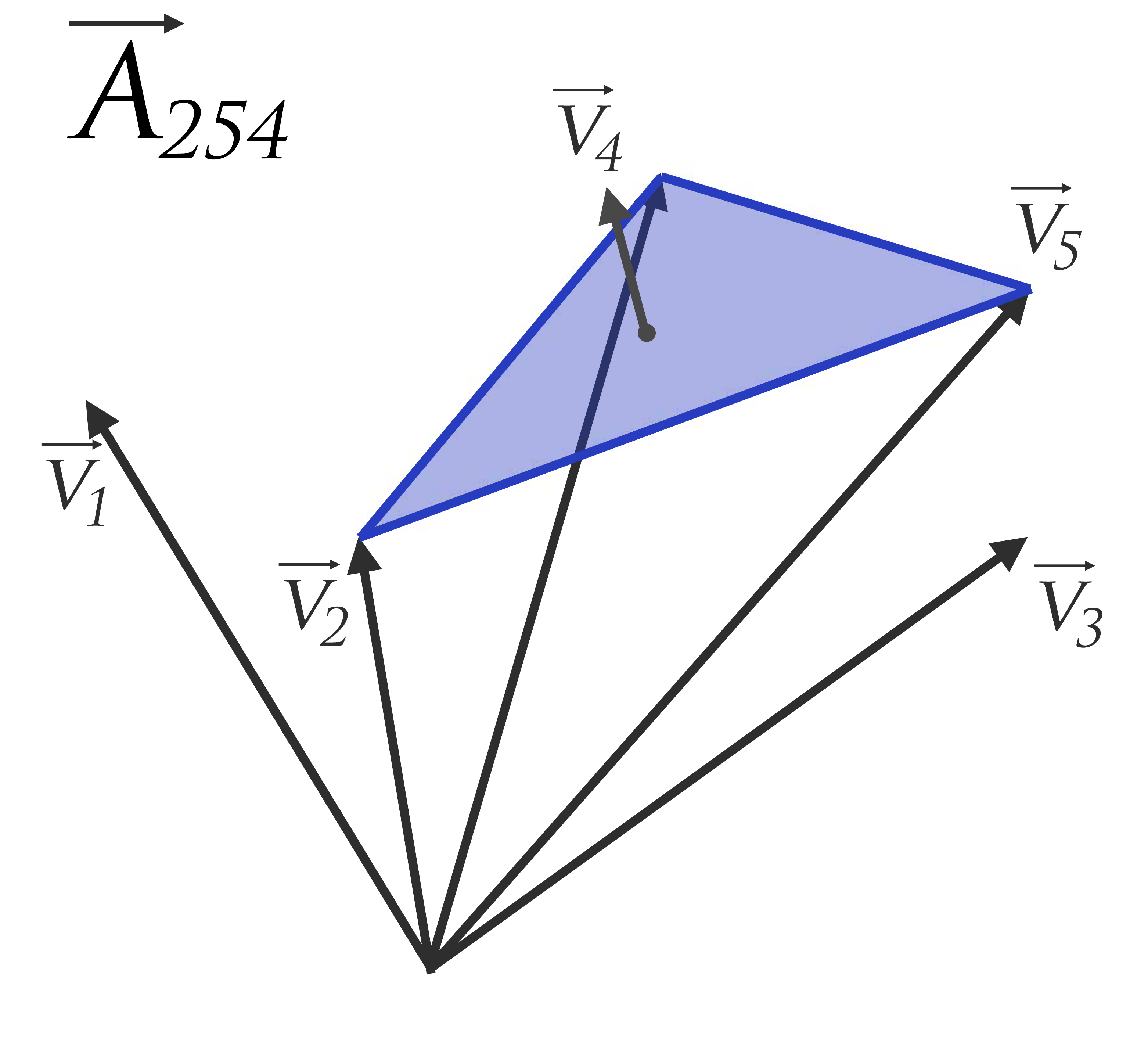}
		\qquad\qquad \includegraphics[width=.2\linewidth]{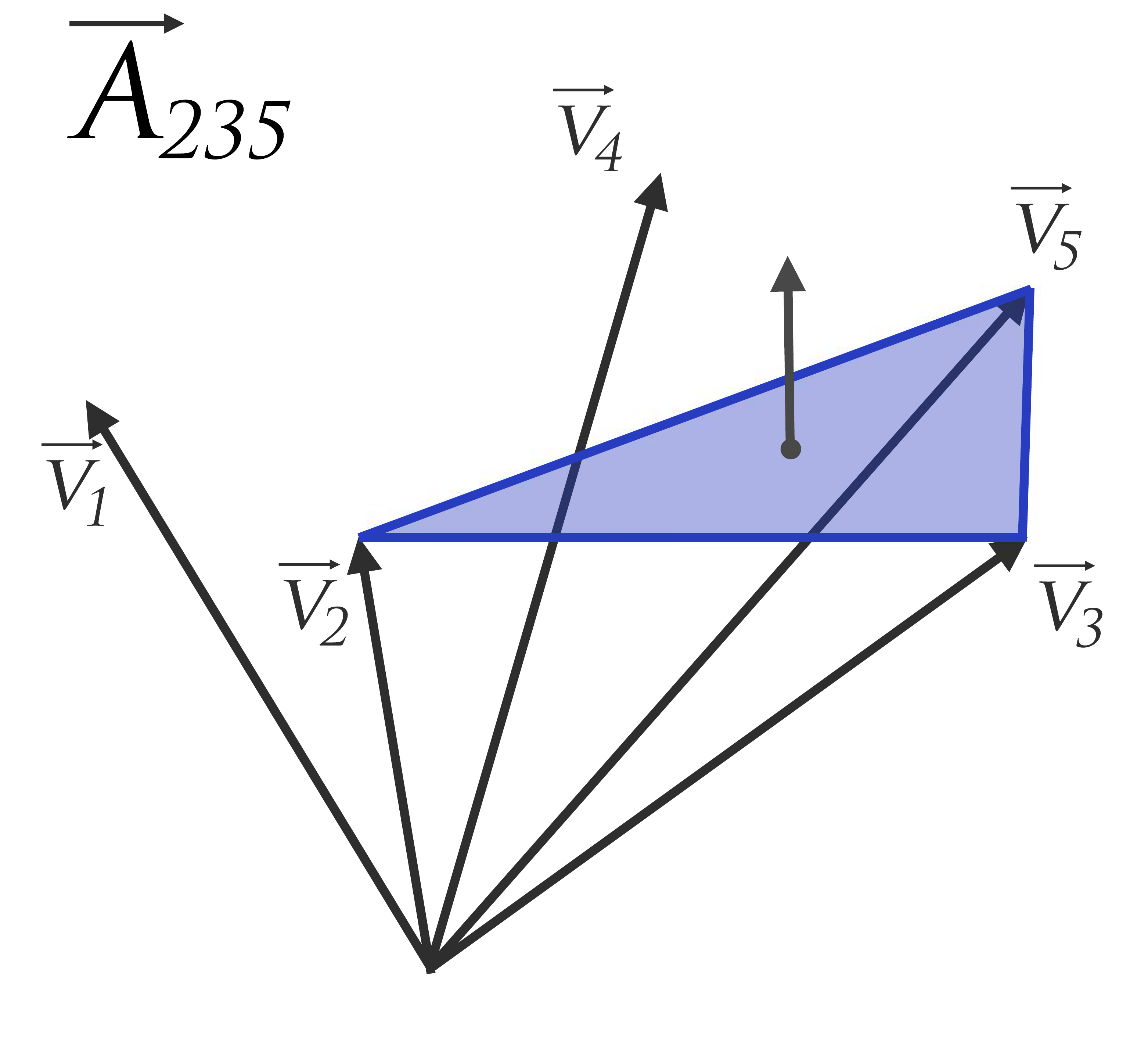}
\caption{Some of the directed areas contributing to (\ref{pinningCurrents}) in $3+1$ dimensions.}
\label{fig:EnergyAreas}
\end{figure}

Similarly to find the currents, select any two condensates $i,j$, and form all $(d-1)$-simplices whose corners are the endpoints of velocity vectors from any $d$ other condensates $i_1, \dots , i_d$. The current is proportional to $\Area^a_{i_1 \cdots i_d}$, weighted by the anti-symmetrized ratios involving the $\mu_i$'s found above. This is illustrated in figure \ref{fig:EnergyAreas}.
Concretely, we have in any dimension
\begin{align}\label{pinningCurrents}
	n_i^A &= 
	\begin{pmatrix}
		\sum\nolimits' f_{i i_1 \cdots i_{d+1}} \frac{1}{m_i} \text{Vol}_{i_1 \cdots i_{d+1}} \\
		\frac{1}{d} \sum\nolimits' f_{ij i_1 \cdots i_d } \left( \frac{1}{m_i} \frac{\mu_j}{m_j} - \frac{1}{m_j} \frac{\mu_i}{m_i} \right) \text{Area}^a_{i_1 \cdots i_d}
	\end{pmatrix}, \nonumber \\
	j^A &= 
	\begin{pmatrix}
		\sum\nolimits' f_{i i_1 \cdots i_{d+1}} \frac{q_i}{m_i} \text{Vol}_{i_1 \cdots i_{d+1}} \\
		\frac{1}{d} \sum\nolimits' f_{ij i_1 \cdots i_d } \left( \frac{q_i}{m_i} \frac{\mu_j}{m_j} - \frac{q_j}{m_j} \frac{\mu_i}{m_i} \right) \text{Area}^a_{i_1 \cdots i_d}
	\end{pmatrix}, \nonumber \\
	\rho^A &= 0, \nonumber \\
	\epsilon^A &= 
	\begin{pmatrix}
		\sum\nolimits' f_{i i_1 \cdots i_{d+1}} \frac{\mu_i}{m_i} \text{Vol}_{i_1 \cdots i_{d+1}} \\
		0
	\end{pmatrix}, \nonumber \\
	T^{ab} &= - \frac{1}{d} \sum\nolimits' f_{ij i_1 \cdots i_d} \left( \frac{\mu_i}{m_i} v^{(a}_j - \frac{\mu_j}{m_j} v^{(a}_i \right) \text{Area}^{b)}_{i_1 \cdots i_d} .
\end{align}

\section{An Example in 1+1 Dimensions}\label{sec:Example}
The utility of an effective action approach is that it bypasses an often intractable microscopic description and allows one to directly write down the most general low energy theory consistent with the symmetries of a problem. However, it is nonetheless instructive to have a microscopic model of the phenomena we've described.
We thus conclude with a simple example of a weakly-coupled model that exhibits nonzero Hall drag. To keep things simple, we will consider the $(1+1)$-dimensional case, where a Hall drag may be obtained in the presence of a bipartite mixture. 

Our characterization of the Hall drag in $1+1$ dimensions is somewhat degenerate since there are no directions perpendicular to the relative velocity. However, the formulas (\ref{hallDragGeneral}) still hold formally with $\Area^a_{i_1} \rightarrow 2$, as one may check from the general expressions (\ref{hallCurrents}). We then see that $c^H$ leads to persistent currents whenever two condensates have overlapping density. In a finite system this will lead to buildup of mass, charge, and energy at the edges of a system with overlapping condensates.

$(1+1)$-dimensional condensation is famously forbidden for translationally invariant systems. However, experimental realizations invariably involve a trapping potential that modifies the density of states sufficiently to circumvent the usual arguments\cite{pethick2002bose} and Bose condensation has been observed in quasi-one-dimensional systems using highly anharmonic traps.\cite{lowDimensionReview} The condensation temperature of the 1-d free Bose gas in such a trap is\cite{pethick2002bose}
\begin{align}
	T_c \approx \omega \frac{N}{\ln N}
\end{align}
where $N$ is the number of atoms and $\omega$ is the trapping frequency in the soft direction. We may formally consider the limit where $N$ is very large and $\omega$ very small with finite $T_c$ and so recover the approximately translationally invariant problem.

\subsection{The Model}\label{sec:Model}

The quasi-one-dimensional problem may be treated as the mean field theory of a three-dimensional system with weak contact interactions in a highly anharmonic trap. It has been argued by variational techniques, supported by numerical evidence, that the effective one-dimensional dynamics is that of a Gross-Pitaevski-like equation with a non-polynomial potential $V(|\psi|^2)$.\cite{salasnich2002effective} 
A similar analysis should yield a non-polynomial $V(|\psi_1|^2, |\psi_2|^2)$ for the quasi-one-dimensional bipartite mixture, but the precise form will not matter for us. The microscopic model we propose is then
\begin{align}\label{model}
	\mathcal L &= \frac{i}{2} \psi_1^\dagger \overset \leftrightarrow \partial_t \psi_1 - \frac{1}{2 m_1} \partial_x \psi_1^\dagger \partial_x \psi_1  + \frac{i}{2} \psi_2^\dagger \overset \leftrightarrow \partial_t \psi_2 - \frac{1}{2 m_2} \partial_x \psi_2^\dagger \partial_x \psi_2 \nonumber \\
		&\qquad \qquad + \frac{i}{2} a \left( m_1 | \psi_1 |^2 \psi_2^\dagger \overset \leftrightarrow \partial_x \psi_2 - m_2 | \psi_2 |^2 \psi^\dagger_1 \overset \leftrightarrow \partial_x \psi_1 \right) 
		-  V ( | \psi_1 |^2 , | \psi_2 |^2 ) .
\end{align}

We have introduced a velocity dependent interaction $a$ that will lead to Hall drag and which is marginal in mean field theory. It is consistent with the symmetries of the problem, as one may see by writting the Lagrangian in a manifestly Galilean invariant form
\begin{align}
	\mathcal L = - \frac{1}{2 m_1} D_I \psi^\dagger_1 D^I \psi_1  - \frac{1}{2 m_2} D_I \psi^\dagger_2 D^I \psi_2 + \frac i 2 a \epsilon_{AB} \psi^\dagger_1 \overset \leftrightarrow D {}^A \psi_1 \psi^\dagger_2 \overset \leftrightarrow D {}^B \psi_2 - V (| \psi_1 |^2 , | \psi_2 |^2 ).
\end{align}
Perhaps a more familiar way of seeing Galilean invariance is to consider the interaction potential between two particles in a single-particle quantum mechanics picture. In a two particle quantum mechanics picture, the field theory (\ref{model}) involves an interaction potential
\begin{align}
	\hat V = \frac 1 2 a m_1 m_2 \left \{ \delta ( \hat X_1 - \hat X_2 ), \frac{1}{m_1} \hat P_1 - \frac{1}{m_2}\hat P_2 \right \} ,
\end{align}
which is Galilean invariant since it references only the relative velocities of the two particles.
While we do not currently have a proposal on how this can be done, it would be interesting to try to engineer such an interaction in future cold atom experiments.

\subsection{Computing the Hall Drag}\label{sec:ModelHallDrag}

In the condensed phase $\psi_i = e^{- i \varphi_i } \sqrt{n_i}$, this reads
\begin{align}
	\mathcal L = n_1 \mu_1 + n_2 \mu_2 + a n_1 n_2 \lambda - V (n_1 , n_2 ) ,
\end{align}
where for this section we are denoting $\mu_i =D_t \varphi_i- \frac{1}{2 m_i} D_x \varphi_i D_x \varphi_i$.
The Gross-Pitaevski equations (GPEs) that follow from varying $n_i$ then read
\begin{align}\label{EOM}
	\mu_1 = \frac{\partial V}{\partial n_1}- a n_2 \lambda ,
	&&\mu_2 =  \frac{\partial V}{\partial n_2}- a n_1 \lambda .
\end{align}
This is the usual form of the equation determining the condensate density in a potential well as a function of the chemical potentials, but now the shape of the well depends on the relative velocities of the condensates.

Plugging in $\mu_1$ and $\mu_2$, we find $\mathcal L$ as a function of the condensate densities and $\lambda$
\begin{align}
	\mathcal L = - a n_1 n_2 \lambda + n_1 \frac{\partial V}{\partial n_1}+ n_2 \frac{\partial V}{\partial n_2}- V .
\end{align}
The pressure is however a function of the chemical potentials $p(\mu_1 , \mu_2 , \lambda)$, so we have
\begin{gather}
	\frac{\partial p}{\partial \lambda} = - a n_1 n_2 - a \frac{ \partial n_1}{\partial \lambda }  n_2 \lambda - a n_1 \frac{ \partial n_2 }{\partial \lambda }  \lambda + n_1 \frac{\partial^2 V}{\partial n_1^2} \frac{ \partial n_1 }{\partial \lambda }  + n_1 \frac{\partial^2 V}{\partial n_1 \partial n_2} \frac{ \partial n_2 }{\partial \lambda } + n_2 \frac{\partial^2 V}{\partial n_1 \partial n_2} \frac{ \partial n_1 }{\partial \lambda }  + n_2 \frac{\partial^2 V}{\partial n_2^2} \frac{ \partial n_2 }{\partial \lambda }  .
\end{gather}
Differentiating the GPEs (\ref{EOM}) with respect to $\lambda$ also gives
\begin{align}
	\frac{\partial^2 V}{\partial n_1^2} \frac{ \partial n_1 }{\partial \lambda }  + \frac{\partial^2 V}{ \partial n_1 \partial n_2} \frac{ \partial n_2 }{\partial \lambda } - a \frac{ \partial n_2 }{\partial \lambda }  \lambda - a n_2 = 0 , \nonumber \\
	\frac{\partial^2 V}{ \partial n_1 \partial n_2} \frac{ \partial n_1 }{\partial \lambda }  + \frac{\partial^2 V}{\partial n_2^2} \frac{ \partial n_2 }{\partial \lambda } - a \frac{ \partial n_1 }{\partial \lambda }  \lambda - a n_1 = 0 .
\end{align}
Plugging these into $\frac{\partial p}{\partial \lambda}$ then gives a Hall drag 
proportional to the product of the mass densities $\rho_i = m_i n_i$
\begin{align}
	c^H = m_1 m_2 \frac{\partial p}{\partial \lambda}= a \rho_1 \rho_2 .
\end{align}

\section{Conclusion}\label{sec:Conclusion}

In this work we have demonstrated a general procedure to construct EFT's for Galilean invariant superfluid mixtures to any order in a momentum expansion. We have carried out the construction to lowest order and found agreement with the results of Greiter, Witten, and Wilczek\cite{Greiter:1989qb} in the single component case. It would also be interesting to look at the next order to confirm agreement with Son and Wingate\cite{Son2006} as well as to investigate what new transport is allowed at this order in the presence of multiple condensates, particularly for atoms at unitarity.

At lowest order, we have found two new parity odd transport coefficients, the Hall drag, and a pinning of mass, charge, and energy to relative velocity, both of which give rise to currents that have simple geometric interpretations in terms of the volumes and areas of various simplices. Both terms require a sufficient number of condensates to be realized that depends on the spatial dimensionality. We have also furnished a weak coupling example in one spatial dimension that exhibits Hall drag. It would be interesting to try to engineer such an interaction in cold atom traps, however, these effects should generically exist in highly dense mixtures with broken $P$ and $T$ and should be observable in experiment.

In this paper we have assumed only Galilean invariance and particle number conservation. However, cold atom mixtures can be created in the lab with a variety of (approximate) flavor symmetries by condensing multiple hyperfine states of a single isotope. See \cite{kasamatsu2005} for a review of these so called ``fictitious spinor'' condensates. It would be interesting to see how these further constrain transport.

True spinor condensates are also experimentally accessible (see \cite{kawaguchi2012spinor} for a review). These systems display a complex phase diagram, realizing different types of magnetic order, and it would be interesting to investigate spin transport in these phases within the formalism we have outlined here.

\begin{acknowledgments}
We are grateful for many fruitful conversations with D. T. Son, M. M. Roberts, and K. Prabhu and input from M. Ueda.
This work is supported by the University of California.
\end{acknowledgments}

\appendix

\section{Superfluid Drag}\label{app:SuperfluidDrag}

In these appendices we collect some of the computational details omitted in our main discussion. The first is that there is an additional set of parity even Galilean scalars that enter at lowest order in derivatives, but were not considered explicitly in the text since their effects are identical to $\mu_{ij}$.

To see this, note that for every pair of fluids, their relative velocity $v^A_i - v^A_j$ is a spatial vector in the sense $(v^A_i - v^A_j)n_A = 0$. Since $h^{AB}$ is $n_A$ orthogonal and furthermore its kernel is spanned by $n_A$, this means that the relative velocity may be written as the ``raised index'' version of a Galilean covector
\begin{align}
	v^A_i - v^A_j = h^{AB} u_{ijB} .
\end{align}
Note this does not uniquely define $u_{ijA}$, as this definition is insensitive to a shift $u_{ijA} \rightarrow u_{ijA} + \xi_{ij} n_A$. The scalar
\begin{align}
	\nu_{ijkl} &= m_i m_j m_k m_l h^{AB} u_{ijA} u_{klB} \nonumber \\
		&= (m_j D_a \varphi_i - m_i D_a \varphi_j )(m_l D^a \varphi_k - m_k D^a \varphi_l)
\end{align}
is however insensitive to this ambiguity and is thus a well-defined Galilean invariant.

The most general lowest-order parity invariant EFT is then
\begin{align}
	S = \int d^{d+1}x | e | p ( \mu_{ij} , \nu_{ijkl} ) .
\end{align}
As in the main text, we define
\begin{align}
	\delta p = \sum_{ij} p_{ij} \delta \mu_{ij} + \sum_{ijkl} p_{ijkl} \delta \nu_{ijkl} ,
\end{align}
where the sums go over all index configurations. Due to the symmetries of $\mu_{ij}$ and $\nu_{ijkl}$, this definition includes some overcounting.

To actually carry out the variations, note that
\begin{align}\label{basicVariations}
	\delta D_I \varphi_i &= - \Pi^\mu{}_I \delta e^J_\mu D_J \varphi_i + q_i \Pi^\mu{}_I \delta A_\mu ,
	&& \implies
	&& \delta D^A \varphi_i = - h^{A\mu} D_I \varphi_i \delta e^I_\mu + q_i h^{A\mu} \delta A_\mu ,
\end{align}
where we have used $\Pi^A{}_I \Pi^{BI} = h^{AB}$.
This implies the variation of $u_{ijA}$ up to a term proportional to $n_A$. Of course, since the invariants $\nu_{ijkl}$ do not depend on the $n_A$ part, we can disregard this ambiguity
\begin{align}\label{uVar}
	\delta u_{ijA} = e^\mu_A \left( \frac{1}{m_i} D_I \varphi_i - \frac{1}{m_j} D_I \varphi_j \right) \delta e^I_\mu - \left( \frac{q_i}{m_i} - \frac{q_j}{m_j} \right) e^\mu_A \delta A_\mu .
\end{align}
We then find that
\begin{align}\label{evenBackgroundVariations}
	\delta \mu_{ij} &= ( D^\mu \varphi_i D_I \varphi_j + D^\mu \varphi_j D_I \varphi_i ) \delta e^I_\mu - ( q_i D^\mu \varphi_j + q_j D^\mu \varphi_i) \delta A_\mu, \nonumber \\
	\delta \nu_{ijkl} &= \left( \left( m_i D^\mu \varphi_j - m_j D^\mu \varphi_i\right) \left( m_l D_I \varphi_k - m_k D_I \varphi_l  \right) + ij \leftrightarrow kl \right) \delta e^I_\mu \nonumber \\
		&\qquad \qquad - \left( \left( m_i D^\mu \varphi_j - m_j D^\mu \varphi_i \right) \left( q_k m_l - q_l m_k \right) + ij \leftrightarrow kl \right) \delta A_\mu .
\end{align}
Using these variations, we find the currents are identical to (\ref{covariantCurrents}-\ref{numberMatrix}), but with a new $S_{ij}$ matrix
\begin{align}\label{sDef}
	S_{ij} = 2 p_{ij} + 8 \sum_{kl} p_{iklj} m_k m_l .
\end{align}

\section{Drag Induced Effective Mass}\label{app:EffectiveMass}
As we have seen, drag leads to the collective motion of the charges of many fluids once a particular fluid has been given some velocity. These collective modes will have a renormalized mass and charge that we can solve for explicitly in terms of the equation of state. To compute these, let's begin by writing (\ref{covariantCurrents}) as
\begin{gather}
	\tau^\mu{}_I = - p \Pi^\mu{}_I -  \sum_{ij} m_i N_{ij} v^\mu_i v_{jI}, \qquad \qquad
	\rho^\mu = \sum_{ij} m_i N_{ij} v^\mu_j , \nonumber \\
	j^\mu = \sum_{ij} q_i N_{ij} v^\mu_j , \qquad \qquad
	n^\mu_i = \sum_j N_{ij} v^\mu_j ,
\end{gather}
where
\begin{align}
	N_{ij} = S_{ij} m_j
\end{align}
and $S_{ij}$ is defined in (\ref{sDef}). Recall $\rho^\mu = - \tau^\mu{}_I n^I$ is the mass current and is Galilean invariant.

For notational simplicity we shall adopt notation where $S$ and $N$ are the matrices with matrix elements $S_{ij}$ and $N_{ij}$ respectively. We also define the column vectors of charges
\begin{align}
	m =
	\begin{pmatrix}
		m_1 \\
		\vdots \\
		m_n
	\end{pmatrix},
	&&
	q =
	\begin{pmatrix}
		q_1 \\
		\vdots \\
		q_n
	\end{pmatrix},
\end{align}
and similarly for the one-index objects $n^\mu$ and $v^\mu$. $M$ denotes the diagonal matrix
\begin{align}
	M = \diag{(m)} = 
	\begin{pmatrix}
		m_1 & & \\
		 & \ddots & \\
		  & & m_n
	\end{pmatrix}.
\end{align}
The currents (\ref{covariantCurrents}) are then
\begin{gather}
	\tau^\mu{}_I = - p \Pi^\mu{}_I -  (v^T)^\mu M N v_I, \qquad \qquad
	\rho^\mu =  m^T N v^\mu,  \nonumber \\
	j^\mu = q^T N v^\mu , \qquad \qquad
	n^\mu = N v^\mu .
\end{gather}

The drag leads to the collective motion of the charges of many fluids once a particular fluid has been given some velocity. The collective modes will then have a renormalized mass and charge that we now solve for explicitly in terms of the equation of state. To do so, diagonalize the symmetric matrix $S$ and define the vector of momentum currents
\begin{align}
	S = O D O^{-1},
	&&p^\mu = M v^\mu.
\end{align}
Now let
\begin{align}
	n^{\star \mu} = O^{-1} n^\mu,
	&&p^{\star \mu} = O^{-1} p^\mu ,
	&&m^\star = O^{-1} m ,
	&&q^\star = O^{-1} q .
\end{align}
As we shall see, the entries of $m^\star$ and $q^\star$ are the masses and charges of the collective modes. If we define $M^\star = \text{diag} ( m^\star )$, $u^\mu = (M^\star)^{-1} p^{\star \mu},$ and $N^\star = D M^\star(\equiv \diag(n^\star_1 , \dots , n^\star_n))$, then the currents take the diagonal form
\begin{gather}
	\tau^\mu{}_I = - p \Pi^\mu{}_I -  (u^T)^\mu M^\star N^\star u_I, \qquad \qquad
	\rho^\mu =  (m^\star)^T N^\star u^\mu, \nonumber \\ 
	j^\mu = (q^\star)^T N^\star u^\mu , \qquad \qquad
	n^{\star \mu} = N^\star u^\mu .
\end{gather}
That is
\begin{gather}
	\tau^\mu{}_I = - p \Pi^\mu{}_I - \sum_i m^\star_i n^\star_i u_i^\mu u_{iI} , \qquad \qquad
	\rho^\mu = \sum_i m^\star_i n^\star_i u^\mu_i , \nonumber \\
	j^\mu = \sum_i q^\star_i n^\star_i u^\mu_i  , \qquad \qquad
	n^{\star \mu}_i = n^\star_i u^\mu_i  .
\end{gather}
$m^\star = O^{-1} m$ and $q^\star = O^{-1} q$ are then the masses and charges of the collective modes $\varphi^\star = O^{-1} \varphi$. In terms of the normal modes, there is no fluid drag.

\section{Parity Odd Transport}\label{app:Parity}
We may similarly define another set of parity odd scalars that encode Hall drag
\begin{align}\label{chi}
	\chi_{i_1 \cdots i_{2d}} &= m_{i_1} \cdots m_{i_{2d}}\epsilon^{A_1 \cdots A_d} u_{i_1 i_2 A_1} \cdots u_{i_{2d-1} i_{2d} A_d} \nonumber \\
		&= \epsilon^{a_1 \cdots a_d} (m_{i_1} D_{a_1} \varphi_{i_2} - m_{i_2} D_{a_1} \varphi_{i_1} ) \cdots (m_{i_{2d-1}} D_{a_d} \varphi_{i_{2d}} - m_{i_{2d}} D_{a_d} \varphi_{i_{2d-1}} ).
\end{align}
Here the epsilon symbol is the spatial epsilon
\begin{align}
	\epsilon^{A_1 \cdots A_d} = n_{A_0 }\epsilon^{A_0 A_1 \cdots A_d}.
\end{align}
The reader should beware as the only notational difference between the spatial and spacetime epsilons is in the number of indices.
Note that one needs $d$ relative velocities to write this down, and hence, like $\lambda_{i_0 \cdots i_d}$, $d+1$ fluids. The parity odd EFT is then
\begin{align}
	S = \int d^{d+1} x | e | p ( \mu_{ij} , \nu_{ijkl}, \lambda_{i_0 \cdots i_d} , \chi_{i_1 \cdots i_{2d}}, \xi_{i_0 \cdots i_{d+1}} ) .
\end{align}

Using (\ref{basicVariations}) and (\ref{uVar}), we find the following variations of the parity odd scalars
\begin{align}\label{oddVariations}
	\delta \lambda_{i_0 \cdots i_d} &= \epsilon^\mu{}_{\mu_1 \cdots \mu_d} \left( D_I \varphi_{i_0} D^{\mu_1} \varphi_{i_1} \cdots D^{\mu_d} \varphi_{i_d} - D_I \varphi_{i_1} D^{\mu_1} \varphi_{i_0} D^{\mu_2} \varphi_{i_2} \cdots D^{\mu_d} \varphi_{i_d} + \cdots \right) \delta e^I_\mu \nonumber \\
		&- \epsilon^\mu{}_{\mu_1 \cdots \mu_d} \left( q_{i_0} D^{\mu_1} \varphi_{i_1} \cdots D^{\mu_d} \varphi_{i_d} - q_{i_1} D^{\mu_1} \varphi_{i_0} D^{\mu_2} \varphi_{i_2} \cdots D^{\mu_d} \varphi_{i_d} + \cdots \right) \delta A_\mu , \nonumber \\
		\delta \chi_{i_1 \cdots i_{2d}} &= m_{i_3} \cdots m_{i_{2d}} \epsilon^{\mu A_2 \cdots A_d}\left( (m_{i_2} D_I \varphi_{i_1} - m_{i_1} D_I \varphi_{i_2} ) u_{i_3 i_4 A_2} \cdots u_{i_{2d-1} i_{2d} A_d}+ \cdots \right) \delta e^I_\mu \nonumber \\
			&-  m_{i_3} \cdots m_{i_{2d}}  \epsilon^{\mu A_2 \cdots A_d} \left(  (m_{i_2} q_{i_1} - m_{i_1} q_{i_2} ) u_{i_3 i_4 A_2} \cdots u_{i_{2d-1} i_{2d} A_d} + \cdots \right) \delta A_\mu, \nonumber \\
	\delta \xi_{i_0 \cdots i_{d+1}} &= -\epsilon^{\mu I_1 \cdots I_{d+1}} \left( D_I \varphi_{i_0} D_{I_1} \varphi_{i_1} \cdots D_{I_{d+1}} \varphi_{i_d} - D_I \varphi_{i_1} D_{I_1} \varphi_{i_0} D_{I_2} \varphi_{i_2} \cdots D_{I_{d+1}} \varphi_{i_{d+1}} + \cdots \right) \delta e^I_\mu \nonumber \\
	&+ \epsilon^{\mu I_1 \cdots I_{d+1}} \left( q_{i_0} D_{I_1} \varphi_{i_1} \cdots D_{I_{d+1}} \varphi_{i_{d+1}} - q_{i_1} D_{I_1} \varphi_{i_0} D_{I_2} \varphi_{i_2} \cdots D_{I_{d+1}} \varphi_{i_{d+1}} + \cdots \right) \delta A_\mu.
\end{align}
Note in the first variation we have the spacetime epsilon with lower indices, the first being raised $\epsilon^\mu{}_{\mu_1 \cdots \mu_d} = h^{\mu \mu_0} \epsilon_{\mu_0 \mu_1\cdots \mu_d}$, in the second we have the spatial epsilon just introduced, while in the final variation we have the extended epsilon symbol $\epsilon^{\mu I_1 \dots I_{d+1} } = \Pi^\mu{}_{I_0} \epsilon^{I_0 I_1 \cdots I_d}$.

Using these, one finds that the currents often include the combination
\begin{align}
		&- (d+1) \sum p_{i i_1 \cdots i_d} \epsilon^A{}_{A_1 \cdots A_d} D^{A_1} \varphi_{i_1} \cdots D^{A_d} \varphi_{i_d} \nonumber \\
		&\qquad \qquad - 2 d \sum p_{i i_2 \cdots i_{2d}} m_{i_2} \cdots m_{i_{2d}} \epsilon^{A A_2 \cdots A_d} u_{i_3 i_4 A_2} \cdots u_{i_{2d-1} i_{2d} A_d} .
\end{align}
Though they look very different, these terms are actually of the same form. To see this, let's expand them out term by term. The temporal part of both is clearly zero. The spatial part of the first term gives
\begin{align}
	- (d+1) \sum p_{i i_1 \cdots i_d} \epsilon^a{}_{A_1 \cdots A_d} D^{A_1} \varphi_{i_1} \cdots D^{A_d} \varphi_{i_d} = - (d+1 ) d \sum p_{i i_1 \cdots i_d} m_{i_1} \epsilon^{a a_2 \cdots a_d} D_{a_2} \varphi_{i_2} \cdots D_{a_d} \varphi_{i_d} ,
\end{align}
while the spatial part of the second is
\begin{align}
	&- 2 d \sum p_{i i_2 \cdots i_{2d}} m_{i_2} \cdots m_{i_{2d}} \epsilon^{a A_2 \cdots A_d} u_{i_3 i_4 A_2} \cdots u_{i_{2d-1} i_{2d} A_d} \nonumber \\
	&\qquad \qquad = - 2 d \sum p_{i i_2 \cdots i_{2d}} m_{i_2} \cdots m_{i_{2d}} \epsilon^{a a_2 \cdots a_d} \left( \frac{1}{m_{i_4}} D_{a_2} \varphi_{i_4} - \frac{1}{m_{i_3}} D_{a_2} \varphi_{i_3}\right) \cdots \left( \frac{1}{m_{i_{2d}}} D_{a_d} \varphi_{i_{2d}} - \frac{1}{m_{i_{2d-1}}} D_{a_d} \varphi_{i_{2d-1}}\right)\nonumber \\
	&\qquad \qquad = - 2 d \sum p_{i i_2 \cdots i_{2d}} m_{i_2} \epsilon^{a a_2 \cdots a_d} \left( m_{i_3} D_{a_2} \varphi_{i_4} - m_{i_4} D_{a_2} \varphi_{i_3}\right) \cdots \left( m_{i_{2d-1}} D_{a_d} \varphi_{i_{2d}} - m_{i_{2d}} D_{a_d} \varphi_{i_{2d-1}}\right)\nonumber \\
	&\qquad \qquad = - 2^d d \sum p_{i i_2 \cdots i_{2d}} m_{i_2} m_{i_3} m_{i_5} \cdots m_{i_{2d-1}} \epsilon^{a a_2 \cdots a_d} D_{a_2} \varphi_{i_4} \cdots  D_{a_d} \varphi_{i_{2d}} .
\end{align}
Adding both together, we have
\begin{align}
	- (d+1) \sum c_{i i_1 \cdots i_d} \epsilon^A{}_{A_1 \cdots A_d} D^{A_1} \varphi_{i_1} \cdots D^{A_d} \varphi_{i_{d+1}}
\end{align}
where
\begin{align}
	c_{i_0 \cdots i_d} = p_{i_0 \cdots i_d} + \frac{2^d}{d+1}\sum_{j_2 \cdots j_d} p_{i_0 i_1 j_2 i_2 \cdots j_d i_d} m_{j_2} \cdots m_{j_d} .
\end{align}

From these, we find that the $\lambda$ and $\chi$ variations lead to the covariant currents
\begin{align}\label{hallCurrents}
	&n_i^\mu = \frac{1}{d!(d-1)!} \epsilon_{\mu_1 \cdots \mu_d}{}^{\mu} \sum_{i_1 \cdots i_d} c^H_{i i_1 \cdots i_d} \frac{1}{m_{i}} v^{\mu_1}_{i_1} \cdots v^{\mu_d}_{i_d} , \nonumber \\
	&j^\mu = \frac{1}{d!(d-1)!} \epsilon_{\mu_1 \cdots \mu_d}{}^{\mu} \sum_{i_0 \cdots i_d} c^H_{i_0 \cdots i_d} \frac{q_{i_0}}{m_{i_0}} v^{\mu_1}_{i_1} \cdots v^{\mu_d}_{i_d} , \nonumber \\
	&\tau^\mu{}_I = - \frac{1}{d!(d-1)!} \epsilon_{\mu_1 \cdots \mu_d}{}^\mu \sum_{i_0 \cdots i_d} c^H_{i_0 \cdots i_d} v_{i_0 I} v^{\mu_1}_{i_1} \cdots v^{\mu_d}_{i_d}  .
\end{align}
where
\begin{align}
	c^H_{i_0 \cdots i_d} = (d+1)! (d-1)! m_{i_0} \cdots m_{i_d} c_{i_0 \cdots i_d}.
\end{align}
We see that as with $\nu_{ijkl}$, the new coefficients $\chi_{i_1 \cdots i_{2d}}$, only alter the formula for the transport coefficients in terms of the equation of state and do not induce transport not already included in the $\lambda_{i_0 \cdots i_d}$'s.

To work the manifestly covariant form (\ref{hallCurrents}) into the component form (\ref{hallDragGeneral}) found in the main text, we need the formula
\begin{align}\label{relativeVelocityIdentity}
	\epsilon_{a a_1 \cdots a_{d-1}} \sum_{i_1 , \dots , i_d } c^H_{i i_1 \cdots i_d} v^{a_1}_{i_1} \cdots v^{a_{d-1}}_{i_{d-1}} = \frac{1}{d} \epsilon_{a a_1 \cdots a_{d-1}} \sum_{i_1 , \dots , i_d } c^H_{i i_1 \cdots i_d}  \Delta v^{a_1}_{i_1 i_2} \Delta v^{a_2}_{i_2 i_3} \cdots \Delta v^{a_{d-1}}_{i_{d-1} i_d} .
\end{align}
To prove this, write
\begin{align}
	&\epsilon_{a a_1 \cdots a_{d-1}} \sum_{i_1 , \dots , i_d } c^H_{i i_1 \cdots i_d} v^{a_1}_{i_1} \cdots v^{a_{d-1}}_{i_{d-1}} 1_{i_d}
		= \epsilon_{a a_1 \cdots a_{d-1}} \sum_{i_1 , \dots , i_d } c^H_{i i_1 \cdots i_d} \Delta v^{a_1}_{i_1 i_2} \Delta v^{a_2}_{i_2 i_3} \cdots \Delta v^{a_{d-2}}_{i_{d-2} i_{d-1}} v^{a_{d-1}}_{i_{d-1} } 1_{i_d} \nonumber \\
		& \qquad \qquad =   \epsilon_{a a_1 \cdots a_{d-1}} \sum_{i_1 , \dots , i_d } c^H_{i i_1 \cdots i_d} \left( \Delta v^{a_1}_{i_1 i_2} \cdots \Delta v^{a_{d-2}}_{i_{d-2} i_{d-1}} \Delta v^{a_{d-1}}_{i_{d-1} i_d } + \Delta v^{a_1}_{i_1 i_2}  \cdots \Delta v^{a_{d-2}}_{i_{d-2} i_{d-1}} v^{a_{d-1}}_{ i_d } \right) \nonumber \\
		& \qquad \qquad =   \epsilon_{a a_1 \cdots a_{d-1}} \sum_{i_1 , \dots , i_d } c^H_{i i_1 \cdots i_d} \left( \Delta v^{a_1}_{i_1 i_2} \cdots \Delta v^{a_{d-2}}_{i_{d-2} i_{d-1}} \Delta v^{a_{d-1}}_{i_{d-1} i_d } - (d-1)v^{a_1}_{i_1 }  \cdots v^{a_{d-1}}_{ i_{d-1}} 1_{ i_d } \right)
\end{align}
where $1_{i_d}$ is the $N$-component object whose entries are all 1 and we have often used the total antisymmetry of $c^H_{i_0 \cdots i_d}$. In the first equality we have used the antisymmetry of $c^H$ to replace velocities with relative velocities. However, we cannot do this for free in the final velocity factor since $v^c_{i_d}$ does not appear anywhere. We perform this replacement and cancel off the additional contribution in the second line. In the third line we expand out the final term and commute around $i_k$ indices using the total antisymmetry of $c^H$. The final term on the RHS is of the same structure as the LHS. We thus obtain (\ref{relativeVelocityIdentity}).

For completeness, we also include the covariant form of the currents induced by the $f$ coefficients
\begin{align}\label{fCovariant}
	&n_i^\mu =    \frac{1}{(d+1)! d!}\epsilon^{\mu I_1 \cdots I_{d+1}} \sum_{i_1 \cdots i_{d+1}} f_{i i_1 \cdots i_d} \frac{1}{m_{i}} (v_{i_1})_{I_1} \cdots (v_{i_{d+1}})_{I_{d+1}} , \nonumber \\
	&j^\mu =    \frac{1}{(d+1)! d!}\epsilon^{\mu I_1 \cdots I_{d+1}} \sum_{i_0 \cdots i_{d+1}} f_{i_0 \cdots i_d} \frac{q_{i_0}}{m_{i_0}} (v_{i_1})_{I_1} \cdots (v_{i_{d+1}})_{I_{d+1}} , \nonumber \\
	&\tau^\mu{}_I = - \frac{1}{(d+1)! d!}\epsilon^{\mu I_1 \cdots I_{d+1}} \sum_{i_0 \cdots i_{d+1}} f_{i_0 \cdots i_{d+1}} (v_{i_0})_I (v_{i_1})_{I_1} \cdots (v_{i_{d+1}})_{I_{d+1}} ,
\end{align}
where
\begin{align}
	f_{i_0 \cdots i_{d+1}} = (-1)^{d+1}(d+2)! d! m_{i_0} \cdots m_{i_{d+1}} p_{i_0 \cdots i_{d+1}} .
\end{align}

\bibliography{SuperfluidEFTBib}

\end{document}